\documentclass[onecolumn]{aastex61}

\usepackage{xcolor}
\usepackage{lineno}
\usepackage{amssymb}
\usepackage{amsmath}

\newdimen\qtwclandw
\newdimen\qtwclandh
\newbox\qtwcfigbox
\newcommand{\blacklandscapefigure}[1]{%
  \clearpage
  \begingroup
    \qtwclandw=\paperheight
    \qtwclandh=\paperwidth
    \ifdefined\pdfpagewidth
      \pdfpagewidth=\qtwclandw
      \pdfpageheight=\qtwclandh
    \fi
    \ifdefined\pagewidth
      \pagewidth=\qtwclandw
      \pageheight=\qtwclandh
    \fi
    \paperwidth=\qtwclandw
    \paperheight=\qtwclandh
    \hoffset=-1in
    \voffset=-1in
    \setbox\qtwcfigbox=\hbox{%
      \rlap{\color{black}\vrule width\qtwclandw height\qtwclandh depth0pt}%
      \vbox to\qtwclandh{%
        \vfil
        \hbox to\qtwclandw{%
          \hfil
          \includegraphics[width=\qtwclandw,height=\qtwclandh,keepaspectratio]{#1}%
          \hfil
        }%
        \vfil
      }%
    }%
    \ht\qtwcfigbox=\qtwclandh
    \dp\qtwcfigbox=0pt
    \wd\qtwcfigbox=\qtwclandw
    \refstepcounter{figure}%
    \label{fig:map}%
    \shipout\box\qtwcfigbox
    \stepcounter{page}%
  \endgroup
}

\shorttitle{Connecting Quarks with the Cosmos: A Quarter-Century Update}

\begin{document}

\title{Ad Astra White Paper:\\ A Pitch for the Next 25 Years of NASA's Physics of the Cosmos Program}

\author[0000-0002-2942-3379]{Eric~Burns}
\affiliation{Department of Physics \& Astronomy, Louisiana State
University, Baton Rouge, LA 70803, USA}
\email{ericburns@lsu.edu}

\author[0000-0001-7863-1126]{Ivan~Agullo}
\affiliation{Instituto de Física Corpuscular (IFIC), CSIC-Universitat de València and Departament de Física Teórica, Facultad de Física, Burjassot-46100, València, Spain}
\affiliation{ValER, Valencian Foundation for Excellence in Research, calle Mayor 83-1º-L9, 12001 Castelló de la Plana, Spain}

\author[0000-0002-8977-1498]{Igor~Andreoni}
\affiliation{Department of Physics and Astronomy, University of North Carolina at Chapel Hill, Chapel Hill, NC 27599-3255, USA}

\author[0009-0004-0656-1850]{Catherine~M.~Deibel}
\affiliation{Department of Physics \& Astronomy, Louisiana State
University, Baton Rouge, LA 70803, USA}

\author[0000-0002-4782-3851]{Thomas~Essinger-Hileman}
\affiliation{NASA Goddard Space Flight Center, 8800 Greenbelt Road, Greenbelt, MD 20771, USA}

\author[0000-0003-2624-0056]{Christopher~L.~Fryer}
\affiliation{Center for Nonlinear Studies, Los Alamos National Laboratory, Los Alamos, NM 87545 USA}

\author[0000-0001-8079-1882]{Natasha~Latouf}
\affiliation{NASA Postdoctoral Program Fellow, NASA Goddard Space Flight Center, 8800 Greenbelt Road, Greenbelt, MD 20771, USA}

\author[0000-0002-2666-728X]{M.~Coleman~Miller}
\affiliation{Department of Astronomy, University of Maryland, College Park, MD 20742, USA}
\affiliation{Joint Space-Science Institute, University of Maryland, College Park, MD 20742, USA}

\author[0000-0002-6548-5622]{Michela~Negro}
\affiliation{Department of Physics \& Astronomy, Louisiana State
University, Baton Rouge, LA 70803, USA}
\email{ericburns@lsu.edu}

\author[0000-0002-9267-6213]{Jillian~C.~Rastinejad}
\altaffiliation{NASA Einstein Fellow}
\affiliation{Department of Astronomy, University of Maryland, College Park, MD~20742, USA}

\author{Breann~N.~Sitarski}
\affiliation{NASA Goddard Space Flight Center, 8800 Greenbelt Road, Greenbelt, MD 20771, USA}

\author[0000-0002-9249-0515]{Zorawar~Wadiasingh}
\affiliation{Department of Astronomy, University of Maryland, College Park, MD~20742, USA}
\affiliation{NASA Goddard Space Flight Center, 8800 Greenbelt Road, Greenbelt, MD 20771, USA}
\affiliation{Center for Research and Exploration in Space Science and Technology, NASA/GSFC, Greenbelt, MD 20771, USA}

\begin{abstract}
Astrophysical observations of our universe have been key to our understanding of how the universe works. Shortly after the turn of the millennium, the National Research Council delivered \textit{Connecting Quarks with the Cosmos: Eleven Science Questions for the New Century}. In the subsequent quarter-century, we have made substantial progress in answering each question. These advancements have, in part, arisen because of the success of major US facilities across several domains of physics, guided by long-term planning documents which still largely focus on these questions. This report seeks to provide a status update on each question, and to outline what space-based facilities are crucial for future progress, intended to guide NASA's preparatory work for the Astro2030 Decadal. 
\end{abstract}

\makeatletter
\@booleanfalse\titlepage@sw
\let\savedtwocolumngrid\twocolumngrid
\let\twocolumngrid\relax
\maketitle
\let\twocolumngrid\savedtwocolumngrid
\global\@firstsectionfalse
\makeatother


\clearpage
\blacklandscapefigure{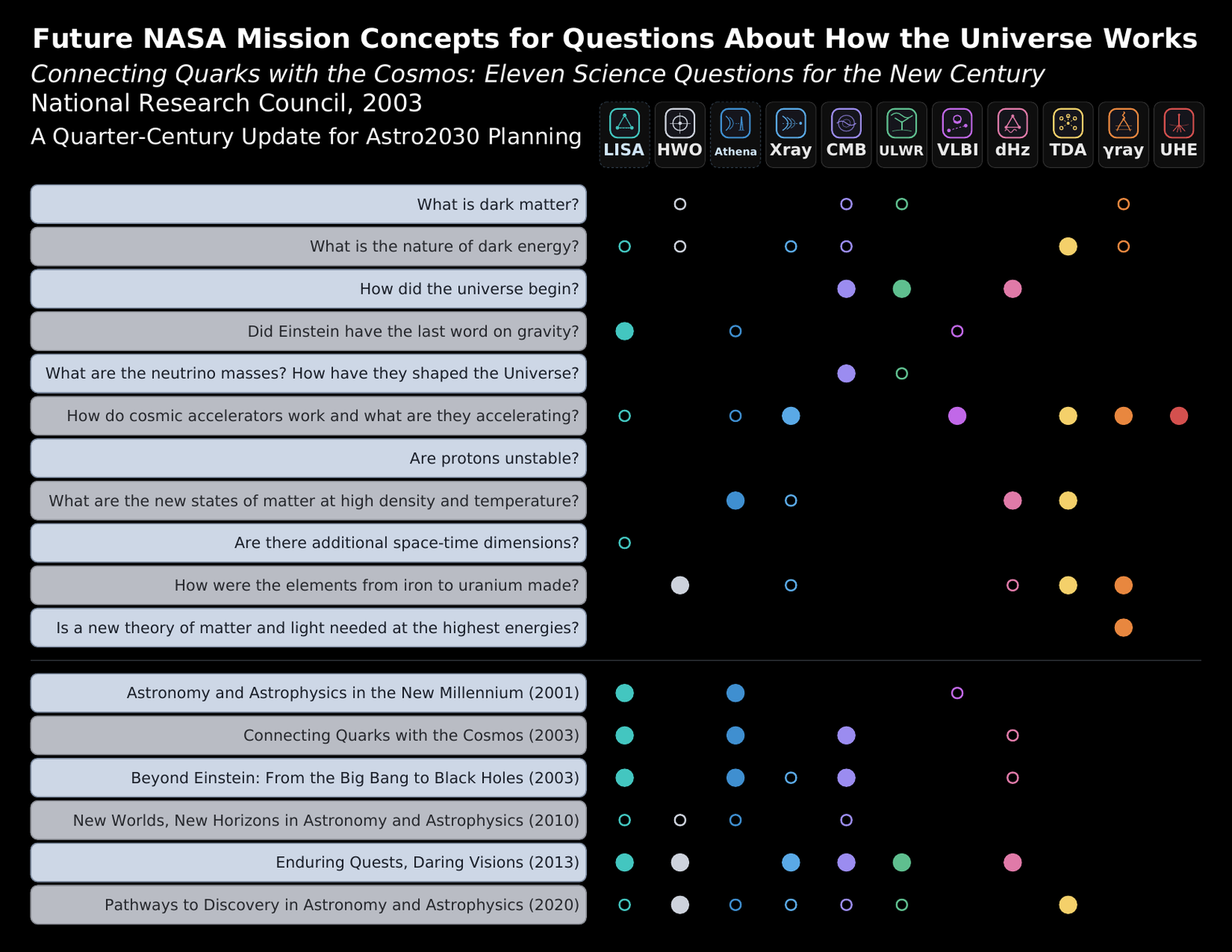}


\clearpage
\blacklandscapefigure{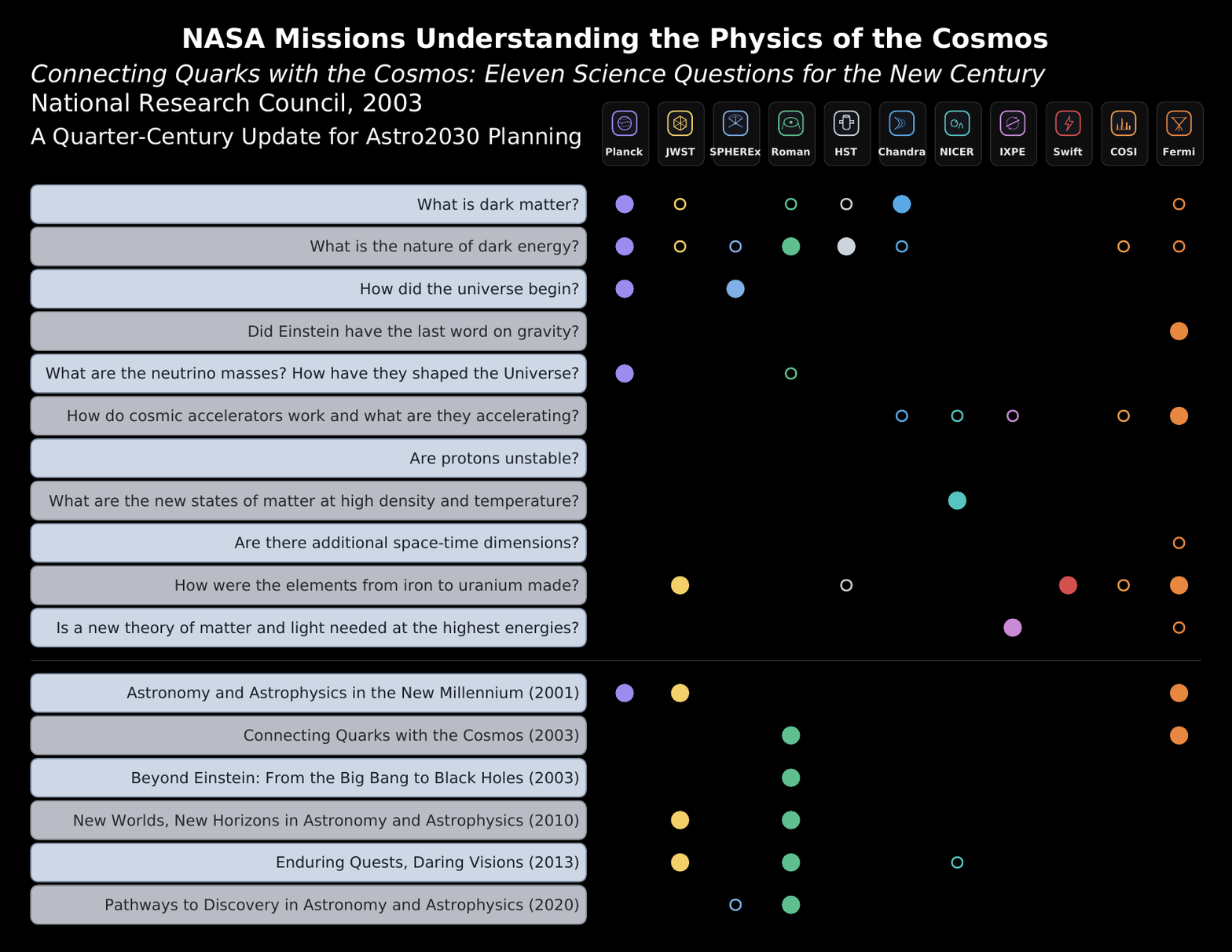}

\clearpage

\pagecolor{white}

\section*{Executive Summary}

Since the Second World War, the United States has led the scientific enterprise of understanding how the universe works. US-based scientists and US-built facilities were central to nearly every discovery that built modern cosmology and high-energy astrophysics, and the record of the field's highest recognitions, the Nobel and Shaw Prizes, reflects that leadership. The ten Nobel Prizes awarded here include  the nuclear reactions that power the stars and form the elements, the detection and precise characterization of the cosmic microwave background (CMB), the theoretical foundations of physical cosmology, the founding of neutrino and X-ray astronomy, the indirect and later the direct detection of gravitational waves, the discovery of the accelerating universe, and the discovery of the supermassive black hole at the center of the Milky Way. This leadership has been guided by strategic national planning and investment, coordinated across NASA, the National Science Foundation (NSF), and the Department of Energy (DOE). 

In 2003, the National Academy, through its Committee on the Physics of the Universe, published \textit{Connecting Quarks with the Cosmos: Eleven Science Questions for the New Century} (Q2C). Its premise was that the nature of matter, energy, space, and time would be answered both by accelerators and by observing the universe. The eleven questions concern dark matter (Q1), dark energy (Q2), the origin of the universe (Q3), the limits of general relativity (Q4), neutrino masses and their imprint on large-scale structure (Q5), cosmic accelerators (Q6), proton stability (Q7), matter at extreme density and temperature (Q8), extra spatial dimensions (Q9), the origin of the heavy elements (Q10), and the completeness of known physics at the highest energies (Q11). These questions have been priorities for DOE, NSF, and NASA in the quarter-century since.  The content of this document is informed by the formal strategic planning documents in Table~\ref{tab:planning}. Guided by them, the United States has invested tens of billions of dollars of public funds in pursuit of the eleven questions, which remain unanswered today.

\begin{deluxetable*}{llll}
\tablecaption{Strategic planning documents informing this report.\label{tab:planning}}
\tablehead{
  \colhead{Year} & \colhead{Abbreviation} & \colhead{Title} & \colhead{Sponsor}
}
\startdata
2001 & AANM & \textit{Astronomy and Astrophysics in the New Millennium} & NASEM \\
2003 & Q2C & \textit{Connecting Quarks with the Cosmos} & NASEM \\
2003 & Beyond Einstein & \textit{Beyond Einstein: From the Big Bang to Black Holes} & NASA \\
2003 & X-Games & \textit{Frontiers in High Energy Density Physics} & NASEM \\
2004 & PoU & \textit{A 21st Century Frontier of Discovery: The Physics of the Universe} & NSTC \\
2004 & Quantum Universe & \textit{Quantum Universe} & HEPAP \\
2004 & NSTC HEDP & \textit{Frontiers for Discovery in High Energy Density Physics} & NSTC \\
2006 & DETF & \textit{Report of the Dark Energy Task Force} & AAAC/HEPAP \\
2006 & EPP-2010 & \textit{Revealing the Hidden Nature of Space and Time} & NASEM \\
2007 & NSAC LRP & \textit{The Frontiers of Nuclear Science: A Long Range Plan} & NSAC \\
2010 & Astro2010 & \textit{New Worlds, New Horizons in Astronomy and Astrophysics} & NASEM \\
2013 & Roadmap & \textit{Enduring Quests, Daring Visions} & NASA \\
2014 & P5 & \textit{Building for Discovery: Strategic Plan for U.S. Particle Physics} & P5 \\
2015 & NSAC LRP & \textit{Reaching for the Horizon: The 2015 Long Range Plan for Nuclear Science} & NSAC \\
2020 & Astro2020 & \textit{Pathways to Discovery in Astronomy and Astrophysics for the 2020s} & NASEM \\
2023 & P5 & \textit{Exploring the Quantum Universe} & P5 \\
2023 & NSAC LRP & \textit{A New Era of Discovery: The 2023 Long Range Plan for Nuclear Science} & NSAC \\
2023 & BPS Decadal & \textit{Thriving in Space: Ensuring the Future of Biological and Physical Sciences Research} & NASEM \\
2024 & MPSAC ngGW & \textit{Next-Generation Gravitational Wave Subcommittee Report} & MPSAC \\
2025 & EPP-2024 & \textit{Elementary Particle Physics: The Higgs and Beyond} & NASEM \\
\enddata
\tablecomments{Sponsor abbreviations: NASEM, National Academies of Sciences, Engineering, and Medicine; NASA, National Aeronautics and Space Administration; NSTC, National Science and Technology Council interagency working group; HEPAP, High Energy Physics Advisory Panel; AAAC, Astronomy and Astrophysics Advisory Committee; NSAC, Nuclear Science Advisory Committee; P5, Particle Physics Project Prioritization Panel; MPSAC, Mathematical and Physical Sciences Advisory Committee (NSF).}\vspace{-0.7 cm}
\end{deluxetable*}

These investments have produced major advances on several of the questions. The direct detection of gravitational waves by LIGO in 2015 created a new branch of astronomy (Q4). The binary neutron-star merger GW170817, observed two years later by LIGO and Fermi and much of NASA's astrophysics fleet, confirmed neutron-star mergers as a site of heavy-element production (Q10), and showed that gravitational waves travel at the speed of light to one part in $10^{15}$ (Q4). This event and the X-ray observations of neutron stars by NICER began constraining the behavior of matter at the densities of neutron-star interiors (Q8). Polarization measurements of the cosmic microwave background with WMAP and Planck have ruled out the simplest models of inflation (Q3), and the direct-detection campaign for dark matter has excluded the canonical candidates over orders of magnitude in interaction strength (Q1). Dark-energy surveys have returned a hint that dark energy may not be constant in time (Q2). IceCube detected high-energy astrophysical neutrinos and associated them with individual sources, and Fermi identified hadronic acceleration in supernova remnants via pion bump detection (Q6). NASA's IXPE telescope provided the strongest evidence yet for vacuum birefringence, a prediction of quantum electrodynamics in a regime inaccessible to any laboratory (Q11).

The committed facilities of the coming decade will further our understanding of several questions. The Nancy Grace Roman Space Telescope, launching this year, joins ESA's Euclid, the NSF-DOE's Vera C. Rubin Observatory, and DOE's Dark Energy Spectroscopic Instrument in a coordinated program designed to determine, at the percent level, whether dark energy evolves (Q2). The Deep Underground Neutrino Experiment, Hyper-Kamiokande, and JUNO will determine the neutrino mass ordering, search for charge-parity violation among neutrinos, and extend the proton-lifetime bound by a factor of several (Q5, Q7). The Facility for Rare Isotope Beams and the planned Electron-Ion Collider anchor the laboratory study of dense matter and of nucleosynthesis (Q8, Q10). ESA's LISA and NewAthena missions, both with NASA contributions and launching in the 2030s, will provide stringent new tests of general relativity (Q4) and measure neutron-star masses and radii at the few-percent level (Q8). The Habitable Worlds Observatory (HWO), Astro2020's recommended next NASA flagship, together with JWST and the Extremely Large Telescopes now under construction, can identify the specific elements formed in neutron-star mergers (Q10).

This report updates the status of each question and is intended to provide input into NASA's preparatory work for the Astro2030 decadal survey. Each section reviews one question, summarizes the progress of the past quarter-century and expectations for committed facilities, and identifies the space-based capabilities that could advance the question over a 5--25 year horizon. Within NASA the eleven questions are pursued chiefly through the Physics of the Cosmos (PhysCOS) Program in the Astrophysics Division. NASA has a role in every question except Proton stability (Q7). For the remainder, further progress depends on measurements from space as well as from the ground. The question-by-question review matches to a fiducial set of NASA mission concepts which could be explored with formal studies, mapped against the eleven questions in the opening Figures and summarized below. We highlight where new understanding is expected ahead of the next Decadal, which may inform concept prioritization.

The nearest-term priority is completion of the program already committed: the surveys of Roman, the NASA flight hardware for LISA, the instrument contribution to NewAthena, and COSI's launch in 2027. COSI's early observations will also sharpen the requirements for several of the concepts that follow.

HWO is in pre-formulation as NASA's next astrophysics flagship. Its physics-of-the-cosmos return depends on design choices being made now. With rapid-response commanding, HWO would capture early kilonovae in the ultraviolet and optical, and its ultraviolet spectroscopy of ancient r-process-enhanced stars would constrain the number of heavy-element sources over Galactic history (Q10). Its angular resolution and stability would permit a two-rung distance ladder for the Hubble constant (Q2) and strong constraints on dark-matter substructure (Q1). All three of these topics are currently explored as Driving Science Cases, and will be considered in the mission design process. The additional gamma-ray instrument under consideration would further contribute to probing the origin of heavy elements (Q10), the nature of dark energy via standard siren cosmology (Q2), understanding the engines of cosmic accelerators (Q6), and probing dense matter (Q8).

A space-based CMB probe has stood in NASA's strategic-mission line since Q2C named it one of three new initiatives, and its science case spans four of the eleven questions. A measurement of primordial B-mode polarization reaching tensor-to-scalar ratio uncertainty level of $\sim10^{-4}$ would either detect the gravitational-wave signature of leading inflationary models or rule them out at high confidence (Q3). The same full-sky maps yield: the cosmic-variance-limited measurement of the reionization optical depth, the parameter that currently caps the cosmological determination of the neutrino masses (Q5); a census of light relic particles from the early universe, along with lensing maps of the dark-matter distribution (Q1); and the early-universe anchor for the final generation of dark-energy surveys (Q2). With CMB-S4 canceled in 2025, the US-based conceptual design of a next generation space Probe is the only viable way to simultaneously make all of these measurements. 

A coordinated time-domain and multimessenger program was Astro2020's highest-priority sustaining activity in space. Its foundation is continuous all-sky gamma-ray monitoring, with rapid X-ray, ultraviolet, and infrared follow-up. Such a fleet would detect and localize neutron-star mergers and magnetar giant flares in which the heaviest elements are made, then characterize each counterpart while it is bright (Q10), constraining the dense-matter equation of state across the merger population in a regime no laboratory can reach (Q8). The same gamma-ray monitors would supply the short gamma-ray bursts that extend the standard-siren Hubble diagram to redshifts of 2--3 (Q2) and provide the electromagnetic counterparts of gravitational-wave and neutrino detections (Q6). 

A next-generation MeV gamma-ray observatory, the successor to COSI, addresses five of the eleven questions. Nuclear-line spectroscopy of supernovae, novae, and kilonovae would measure isotopic yields, ejecta velocities, and explosion geometries, identifying where the isotopes are made (Q10). Applied to large samples of Type Ia supernovae, the same lines could place the standardization of the dark-energy Hubble diagram on a physical rather than empirical footing (Q2). Greater continuum sensitivity and angular resolution between 0.1 MeV and a GeV would test for the predicted 1--10 MeV counterparts of the hidden neutrino sources and trace the pion-decay signature of proton acceleration across the Galactic accelerator population (Q6). Polarimetry of magnetars would permit the first tests of photon splitting, extending the study of the strong-field quantum vacuum begun by IXPE (Q11). At the upper end of the band, determining the origin of the Galactic Center Excess, the leading unresolved candidate signal of dark-matter annihilation, may require angular resolution at GeV energies well beyond Fermi's (Q1). The strength of the gamma-ray cases for Q1 and Q6 are depending on the results of COSI, IceCube and the Cherenkov Telescope Array Observatory results in the coming years. The combined requirements of line sensitivity, continuum sensitivity, angular resolution, and polarimetric response likely requires a gamma-ray mission to have multiple instruments, analogous to the Compton Gamma-Ray Observatory. 

A next-generation X-ray observatory beyond NewAthena could characterize the coronae in which the hidden neutrino sources may accelerate particles (Q6), while extending neutron-star mass and radius measurements to a much larger sample (Q8), probing thermonuclear bursts for nuclear astrophysics (Q10), and, could study the quantum vacuum around magnetars (Q10). The importance for X-rays to answering Q6 will clarify in the coming years. A flagship-level study could establish which capabilities, among large collecting area, high spectral resolution, hard-X-ray coverage, polarimetry, and timing, are the most important. It can also explore fulfilling these capabilities with multiple instruments or satellites. The X-ray interferometer of the Roadmap's Visionary Era, able to image the jet-launching region directly, remains the long-term aim of this line.

A radio array on the far side of the Moon would measure the 21 cm signal from the cosmic dark ages, the time between recombination and the first stars. The measurement would extend the primordial power spectrum several decades in scale beyond the reach of the CMB, with sensitivity to primordial non-Gaussianity beyond planned galaxy surveys (Q3), and would trace the clustering and temperature of matter before the first stars formed, and is sensitive to dark-matter interactions and energy injection (Q1). A mature array would ultimately precisely measure the sum of the neutrino masses through the free-streaming suppression of small-scale structure (Q5). The LuSEE-Night pathfinder is scheduled to fly in 2027, tying this concept to the lunar exploration program.

A decihertz gravitational-wave observatory would occupy the band between LISA and the ground-based detectors. There it would sample the primordial gravitational-wave spectrum roughly sixteen orders of magnitude in frequency above the CMB, constraining inflation and reheating at times no other observation can reach (Q3). It would also detect neutron-star binaries days before they merge and localize them in advance, so that their explosions could be observed across the electromagnetic spectrum from onset (Q10). 

A set of space-based observatories viewing the atmosphere from above would collect the highest-energy cosmic rays and neutrinos with uniform exposure over the full sky, at the energies where magnetic deflection may be small enough for charged-particle astronomy to point to sources (Q6). It is the space-based successor to the Auger recommendation with which the Q2C program began.

The findings above are consistent with those from the Physics of the Cosmos Science Analysis Group (SAG) reports, as well as the reports from the series of Time-Domain and Multimessenger Astrophysics workshops. These include the Gamma-ray Transient Network SAG and the Future Innovations in Gamma Rays SAG. The nuclear astrophysics requirements will be detailed by the Isotopes SAG.

\clearpage
\tableofcontents
\clearpage

\pagebreak
\section{Q1: What Is Dark Matter?}

What we know about dark matter comes from astrophysical observations of its gravity: the motions of galaxies within clusters \citep{1933AcHPh...6..110Z}, the flat rotation curves of spiral galaxies \citep{1970ApJ...159..379R}, the acoustic peaks of the cosmic microwave background (CMB), gravitational lensing maps, and the spatial offset between mass and gas in the Bullet Cluster \citep{2006ApJ...648L.109C}. Observations imply that dark matter is non-baryonic, cold (non-relativistic) on cosmological scales, and effectively collisionless on the scales so far observed, and that it makes up 27\% of the mass-energy of the universe \citep{2020A&A...641A...6P}. It has never been conclusively observed to emit, absorb, or reflect light. Searches proceed by direct detection underground, production at colliders, and astrophysical signatures of interactions and of gravitational clustering.

A quarter-century of increasingly sensitive searches has not produced a detection, and the considered mass range now spans more than forty orders of magnitude. The most recent particle-physics consensus study, EPP-2024, still ranks two single-particle candidates first: the quantum chromodynamics (QCD) axion at the light end of that range and the weakly interacting massive particle (WIMP) at the heavy end \citep{medicine2025elementary}. 

\medskip
\noindent\textbf{Dark Sector.} In the early universe, dark matter was spread nearly uniformly, with slight variations in density from place to place. Gravity amplified those variations: regions slightly denser than average drew in surrounding matter and collapsed into halos, self-gravitating clumps of dark matter within which gas could later gather and form galaxies. In cold dark matter this growth runs from the bottom up, the smallest halos collapsing first and merging into ever larger ones, and its outcome is summarized by the halo mass function, the abundance of halos per unit mass \citep{1974ApJ...187..425P}. Observations have confirmed its form from galaxy clusters down to the halos of the faintest galaxies, a span of roughly seven orders of magnitude. The prediction extends much further: the mass function continues as an unbroken power law down to a cutoff set by the microphysics of the dark-matter particle itself, the scale below which its motion in the early universe erased the primordial fluctuations. For the benchmark 100 GeV WIMP that cutoff lies near an Earth mass; warmer or wave-like candidates would raise it by orders of magnitude, into the observable regime. The low-mass end of the halo mass function is directly tied to properties of the dark matter particles.

Within a galaxy halo, the small halos that fall in and survive persist as subhalos, the smallest containing no stars and the larger ones hosting the ultra-faint dwarf galaxies. Any measured deviation from the cold collisionless prediction, a raised cutoff, a self-interaction cross-section, or wave-like behavior of an ultralight candidate, would imply physics beyond a single minimal particle. EPP-2024 describes this as a dark sector with its own particles and forces \citep{medicine2025elementary}, a question Astro2020 carried within New Messengers and New Physics \citep{2021pdaa.book.....N}.

Substructure observations test this directly, and the past quarter-century has tightened the constraints substantially. Wide-field surveys have grown the census of Milky Way satellites from roughly ten in 2003 to more than sixty confirmed or candidate systems \citep{2025OJAp....8E.142P}, the faintest of them among the most dark-matter-dominated objects known. Strong lensing with the Hubble Space Telescope (HST) and Keck probes the subhalo mass function below the mass of any visible galaxy, and combined with the satellite census excludes thermal-relic warm dark matter below roughly 10 keV \citep{2021ApJ...917....7N}. Gaia astrometry has revealed gaps in stellar streams consistent with perturbations by dark subhalos \citep{2019ApJ...880...38B}. Pulsar timing could eventually reach far smaller masses. A clump passing near a millisecond pulsar, or near the Earth, shifts the pulse arrival times, and timing arrays are forecast to sense individual clumps down to roughly Earth mass, the WIMP cutoff scale, once the Square Kilometre Array (SKA) has timed hundreds of pulsars for decades \citep{2019PhRvD.100b3003D}.  Pulsar timing, and timing of eclipsing binaries, can also yield acceleration measurements in the Galaxy, and anomalous accelerations relative to the smooth component of the Galaxy can yield information about subhalo masses and locations \citep{2026arXiv260631042D}. 

On cluster scales, mergers in which X-ray imaging locates the gas and lensing locates the mass limit the self-interaction cross-section to of order a square centimeter per gram \citep{2008ApJ...679.1173R, 2015Sci...347.1462H}. Two other candidates are now tightly constrained: microlensing, gamma-ray, and CMB data allow primordial black holes to be the whole of the dark matter only in the asteroid-mass window \citep{2021RPPh...84k6902C}, and X-ray searches have reduced the allowed sterile-neutrino parameter space, with XRISM finding no sign of the contested 3.5 keV line \citep{2025ApJ...994L..28X}. The measurements remain consistent with a single cold collisionless species.

Forthcoming surveys will go further. Rubin is forecast to roughly double the satellite census and to find of order 100,000 strong lenses, the parent sample for high-resolution follow-up \citep{2025OJAp....8E..89T,2015ApJ...811...20C}. Roman microlensing would reach the remaining primordial-black-hole window \citep{2023PDU....4101231B}.


The Habitable Worlds Observatory (HWO) would extend the substructure measurements beyond these surveys. Its angular resolution and image stability would sharpen strong- and weak-lensing constraints on the subhalo mass function \citep{he2026habitable}, and a dedicated high-precision astrometric instrument would measure the stellar motions that the substructure induces in dwarf galaxies and streams, constraining the self-interaction cross-section and the mass of wave-like candidates \citep{2026ASPC..543...65M}. After reionization, baryonic feedback affects small-scale structure and has to be modeled. 

\begin{quote}
\textit{What could NASA do?} The smallest dark-matter structures in the present-day universe are a good test of whether dark matter is a single particle or part of a richer dark sector with its own particles and forces. HWO with a dedicated astrometric instrument would extend lensing constraints and measure the stellar motions that substructure induces in dwarfs and streams.
\end{quote}

Observables from before the first stars formed are free of baryonic feedback. These can be probed with a future space-based CMB Probe. The effective number of relativistic species, $N_{\rm eff}$, counts every light species that was ever in thermal contact with the primordial plasma: any new particle that reached equilibrium with the Standard Model raises $N_{\rm eff}$ by at least 0.027, a floor set by entropy conservation \citep{2016arXiv161002743A}. Planck plus baryon acoustic oscillation (BAO) data still sit well above this floor, with an uncertainty of 0.17 \citep{2020A&A...641A...6P}. A measurement several times more precise, the goal of the next generation, would approach the floor itself, a test of dark-sector thermal history unavailable from present-day structure. The same maps would reconstruct the projected matter distribution through gravitational lensing of the CMB over the full sky, substantially tightening constraints on the small-scale suppression from warm, wave-like, or baryon-scattering candidates, and would sharpen the recombination-era ionization bound on annihilating dark matter, a channel now within a factor of two of its cosmic-variance limit, which a wide-sky satellite paired with ground data would saturate \citep{2026PhRvD.113l3018M}.

\begin{quote}
\textit{What could NASA do?} A future space-based CMB Probe would test dark matter in the early universe. It would count any light particle that was ever in thermal contact with ordinary matter, bound dark-matter annihilation (including the candidate signal from Fermi's observations of the Galactic Center), and probe primordial small-scale structure, together seeking evidence of a dark sector.
\end{quote}

This can also be probed by detecting the dark ages 21 cm signal, observable only from the lunar far side. In the dark ages the neutral hydrogen traces the matter distribution directly, and the signal reaches far finer scales than the CMB itself: the CMB is smoothed on small scales by photon diffusion in the primordial plasma, while the 21 cm fluctuations survive down to the much smaller scale set by gas pressure, and they fill a three-dimensional volume rather than a single surface, multiplying the number of independent modes by orders of magnitude \citep{2004PhRvL..92u1301L}. This enables the measurement of small scales in the matter power spectrum, still in the linear regime, where free streaming, wave-like behavior, and dark-sector interactions are least mixed with later astrophysics. The gas is also the coldest ordinary matter in cosmic history, so dark-matter scattering would cool it further, while annihilation, decay, or primordial black holes would heat it \citep{2018Natur.557..684M}.

\begin{quote}
\textit{What could NASA do?} Ultra-long wavelength radio via a lunar far-side 21 cm array would map small-scale dark-matter clustering before the first stars, including scattering, annihilation, decay, or primordial-black-hole heating of the gas.
\end{quote}

Neutron stars offer a different test. Dark matter bound inside them would alter their mass-radius relation, adding an intrinsic scatter that could track where the stars formed \citep{2023PhRvD.107j3051R, 2025PhRvD.111l3034R}. Capture over a star's lifetime supplies far too little mass for a measurable effect, so any such component must be present at formation. The large sample of precise mass and radius measurements sought for Q8, anchored by laboratory constraints on the baryonic equation of state, would test for this scatter.

\medskip
\noindent\textbf{Axions.} Charge-Parity (CP) symmetry requires the laws of physics to be unchanged when every particle is replaced by its antiparticle and the universe is viewed in mirror reflection. The weak interaction is known to violate this symmetry; the strong interaction, as far as any measurement can tell, does not, even though its equations permit a term that would. This is the strong CP problem. The neutron electric dipole moment requires that coefficient to be smaller than $10^{-10}$, with no Standard Model reason for it to be small. The Peccei--Quinn mechanism provides a solution, but requires the QCD axion to drive that term to zero dynamically \citep{1977PhRvL..38.1440P}. When the associated symmetry breaks in the early universe, the axion field is left displaced from the minimum of its potential; once the universe cools and the axion acquires its mass, the field oscillates, and the energy of those oscillations dilutes exactly as cold matter does \citep{1983PhLB..120..127P}. The axion would be dark matter as a coherent field filling space, cold because it is born essentially at rest.

Over much of its natural mass range, roughly a neV to a meV, that energy matches the observed dark-matter abundance. Heavier QCD axions are excluded by astrophysics. Because every coupling of the axion scales with its mass, a limit on its coupling to nucleons is a limit on its mass. Axion emission would have shortened the neutrino burst of SN 1987A and sped the cooling of isolated neutron stars measured in X-rays unless the mass lies below roughly 15 meV \citep{2019JCAP...10..016C, 2022PhRvL.128i1102B}. Within the natural window, the search is terrestrial. Microwave-cavity haloscopes convert axions to photons in strong magnetic fields and have reached the theoretically motivated coupling over part of this range \citep{2025PhRvL.134k1002G}. This effort could discover the QCD axion or close the favored window by the late 2030s. No foreseeable space platform approaches this sensitivity.



\medskip
\noindent\textbf{WIMPs.} The hierarchy problem is the unexplained lightness of the Higgs boson. Quantum corrections pull its mass toward the highest energy scales, seventeen orders of magnitude above the measured value, unless new particles near the weak scale cancel that pull. Supersymmetry and the other frameworks built to supply those particles generically contain a stable one at the GeV--TeV scale, the WIMP. Such a particle would have been produced thermally in the hot early universe: created and destroyed in equilibrium with ordinary matter until the expansion diluted the universe enough that annihilations could no longer keep pace, freezing out a relic population. For weak-scale mass and weak-scale interactions, the surviving abundance comes out close to the observed dark-matter density, an alignment of independent arguments known as the ``WIMP miracle.'' The natural window runs from roughly a GeV at the Lee-Weinberg lower bound, below which a thermal relic would overproduce dark matter \citep{1977PhRvL..39..165L}, up to roughly 100 TeV at the unitarity upper bound, above which the required annihilation cross-section becomes inconsistent with the observed abundance. Direct detection, gamma-ray telescopes, and air-shower arrays excluded portions of this window.

In the GeV--TeV window, liquid-xenon time-projection chambers led by LUX-ZEPLIN (LZ) have tightened the upper limit on the spin-independent WIMP-nucleon cross-section by nearly six orders of magnitude since 2003, removing the canonical supersymmetric parameter space over most of the mass range \citep{2025PhRvL.135a1802A}. The program has reached the neutrino fog to a few GeV, the irreducible background from coherent neutrino-nucleus scattering \citep{lz2025solar}. XLZD, the generation-3 successor prioritized by the 2023 P5 report, is designed to reach the fog through the core of the GeV--TeV window \citep{2025EPJC...85.1192X}, though it only probes nuclear scattering. DOE has paused U.S. support for a next-generation xenon experiment. 

In the same mass range, the Large Area Telescope on the Fermi Gamma-ray Space Telescope (Fermi-LAT) tests WIMP annihilation. Dwarf-galaxy observations place the annihilation limit near or below the thermal-relic value at tens of GeV \citep{2020PhRvD.102f1302A,2024PhRvD.109f3024M}. The Galactic Center GeV Excess (GCE) remains the leading unresolved candidate signal: extended emission toward the inner Galaxy whose spectrum and near-spherical morphology are consistent with a 30--50 GeV WIMP annihilating in a cuspy halo \citep{2009arXiv0910.2998G}. However, it is also consistent with an unrecognized population of millisecond pulsars, or with mismodeled Galactic diffuse emission \citep{2009arXiv0910.2998G,2020ARNPS..70..455M,list2026gce}. Only a few millisecond pulsars are known in the bulge, where dispersion and scattering defeat most timing searches, but an arcsecond radio survey of three square degrees there found about a hundred steep-spectrum point sources, the number of pulsars expected at its depth, where five were catalogued \citep{2024MNRAS.531.2191P}.

The Cherenkov Telescope Array Observatory (CTAO), with construction under way at both sites and early science expected within the next few years, will probe annihilation from roughly 100 GeV to 10 TeV with sensitivity below the thermal-relic cross-section \citep{2021JCAP...01..057A}, and will test the millisecond-pulsar hypothesis for the GCE through the TeV inverse-Compton emission such a population must produce. The 10--100 TeV interval up to the unitarity bound requires a southern wide-field air-shower observatory viewing the Galactic Center. The Southern Wide-field Gamma-ray Observatory (SWGO) is the leading concept, though not yet funded \citep{2025arXiv250601786S}. This regime is accessible only from the ground.

\begin{quote}
\textit{What could NASA do?} The Galactic Center Excess, a persistent gamma-ray glow toward the inner Milky Way, may be a signature of dark-matter particles annihilating. Resolving the Galactic Center Excess requires better GeV angular resolution, accessible only from space. CTAO, and possibly SWGO, will look for dark matter annihilation at higher energies. These facilities and ground-based deep radio searches will seek Galactic center pulsars to explore the leading alternative. Thus, the importance of a future space-based gamma-ray mission for this signal will be informed with results ahead of 2030.
\end{quote}

Below the GeV scale, one older candidate signal is still open. INTEGRAL has mapped 511 keV positron-annihilation emission concentrated in the Galactic bulge, which requires some $10^{43}$ positrons per second from sources not yet identified \citep{2005A&A...441..513K, 2016A&A...586A..84S}. Dark matter of a few MeV annihilating to electron-positron pairs was an early proposal \citep{2004PhRvL..92j1301B}. The absence of in-flight annihilation emission limits the positron injection energy to a few MeV \citep{2006PhRvL..97g1102B}, and the CMB excludes the simplest thermal versions \citep{2020A&A...641A...6P}, but the source population remains unsettled. The Compton Spectrometer and Imager (COSI) will image the emission to identify the positron sources.

\section{Q2: What Is the Nature of Dark Energy?}

Dark energy is 68\% of the present mass-energy \citep{2020A&A...641A...6P} and has never been detected through any interaction except gravity. Among the known components of the present universe, it is the only one whose gravity drives the expansion to accelerate rather than decelerate. Its existence was discovered in 1998 by two independent teams using Type Ia supernovae as standard candles, who found that distant supernovae are dimmer than they would be in a decelerating universe \citep{1998AJ....116.1009R, 1999ApJ...517..565P}. The simplest explanation is the cosmological constant $\Lambda$ that Einstein introduced in 1917, an irreducible energy density of empty space whose equation of state $w = P/\rho = -1$ does not evolve \citep{1917SPAW.......142E}. Quantum field theory does not predict this energy density. Naive vacuum-energy estimates miss the measured dark matter energy density by up to 120 orders of magnitude.

\medskip
\noindent\textbf{Constant or Evolving?} Two complementary programs test whether $w$ departs from $-1$. Expansion-history measurements (Type Ia supernovae, the baryon acoustic oscillation (BAO) feature in galaxy clustering, and the present-day expansion rate $H_0$) trace the distance-redshift relation directly; growth-of-structure measurements (weak gravitational lensing, redshift-space distortions, cluster abundance) track how matter clumps under gravity, which dark energy slows. Galaxy clusters contribute to both, as their abundance as a function of mass and redshift measures growth, and the fraction of their mass in hot gas, which X-ray observations measure, traces the distance-redshift relation \citep{2004MNRAS.353..457A, 2008MNRAS.383..879A}. Counting clusters means finding them, through the Sunyaev-Zeldovich decrement in cosmic microwave background (CMB) maps or X-ray surveys, and then characterizing them with X-ray observations \citep{2011ARA&A..49..409A}.

The multi-facility expansion and growth programs are the longest-running example of a coherent NASA, NSF, and DOE effort to answer a single question. Q2C named a wide-field dark-energy space telescope one of its three new initiatives. The 2006 DETF report mapped understanding the equation of state of dark energy into a sequence of staged experiments across NASA, DOE, and NSF, and successive decadal surveys and P5 reports carried the facility sequence through prioritization, funding, and construction. Each completed stage reinforced the cosmological constant: WMAP and Planck pinned the dark-energy density to about 1\%, the Sloan Digital Sky Survey (SDSS) detected the BAO feature in 2005 \citep{2005ApJ...633..560E}, and the supernova Hubble diagram grew through Pantheon+ and the five-year Dark Energy Survey sample \citep{2022ApJ...938..110B, 2024ApJ...973L..14D}.

That consistency is now challenged. The Dark Energy Spectroscopic Instrument (DESI) reported a preference for an evolving equation of state, phantom-like ($w < -1$) at higher redshift and crossing to quintessence-like values ($w > -1$) today. Combined with CMB data, the preference over the cosmological constant, cold dark matter ($\Lambda$CDM) model is 3.1$\sigma$ from BAO and between roughly 3 and 4$\sigma$ when a supernova sample is added, depending on the compilation \citep{2025PhRvD.112h3515A}. Two years of scrutiny have not resolved this disagreement: cross-calibrated supernova reanalyses reduce the preference of $\Lambda$CDM without eliminating it, while analyses that question the consistency of the BAO and supernova data, or that allow a higher CMB optical depth to reionization $\tau$, can remove it \citep{2026MNRAS.548ag632P, afroz2026hint, 2026arXiv260630903S}. Past anomalies have resolved on very different timescales: the 2014 BICEP2 B-mode signal was traced to Galactic dust within a year, while the $S_8$ discrepancy between weak-lensing and CMB clustering amplitudes took a decade to ease and has not fully closed.

A confirmed evolution would falsify the cosmological constant that anchors $\Lambda$CDM and leave two broad classes of explanation: a dynamical field or a dark sector with additional degrees of freedom evolving within General Relativity, or a modification of the gravitational field equations on cosmological scales. Distinguishing between them would then be the main cosmological problem, and either answer would be new fundamental physics.

The final stage of that sequence is now operating or imminent. DESI's five-year analysis will be the next major result released. Euclid is surveying for weak lensing and galaxy clustering. Rubin will produce a deeper weak-lensing map than any existing survey and a Type Ia sample larger than all previous surveys combined. Roman launches this year; its high-latitude surveys deliver weak lensing, BAO and redshift-space distortions, and the first large Type Ia sample at $z > 1$, where evolving and constant models diverge most \citep{2015arXiv150303757S, 2021arXiv211103081R}. Spec-S5, the DESI successor prioritized by the 2023 P5 report, would extend the BAO measurement to still higher redshift \citep{2025arXiv250307923B}. Together they are designed to determine whether dark energy is a cosmological constant at the percent level.

If evolution is confirmed, those surveys start to measure $w(z)$. If $\Lambda$ survives, the tighter limits are the target a vacuum theory must meet. Concordance precision is then set at two ends. On the early side, uncertainty in $\tau$ is among the dominant degeneracies in the joint fit, and an upward shift can remove the DESI preference. Only wide-area polarization mapping from space reaches the cosmic-variance limit on $\tau$, which a CMB Probe would provide. The corresponding late-universe anchor is $H_0$.

\begin{quote}
\textit{What could NASA do?} Concordance cosmology uses data from every epoch of the universe. The final Roman, Rubin, Euclid, and Spec-S5 surveys will transform our understanding of the late universe; a space CMB Probe could do the same for the early-universe data, including a cosmic-variance-limited $\tau$.
\end{quote}

\medskip
\noindent\textbf{The Hubble Constant.} $H_0$ is the cumulative effect of every component of cosmic energy density over 13.8 billion years. It sets the scale for present-day distances, ages, and densities, and joint cosmological fits must reproduce its measured value. The usual local determination is a distance ladder: geometric anchors calibrate stellar indicators in nearby galaxies, the stellar indicators calibrate Type Ia supernovae, and the supernovae carry the measurement into the smooth Hubble flow, where cosmic expansion dominates over peculiar motion. The anchors are Gaia parallaxes, detached eclipsing binaries in the Large Magellanic Cloud, which fix its distance to one percent \citep{2019Natur.567..200P}, and the water-maser disk of NGC 4258 \citep{2019ApJ...886L..27R}. Each rung inherits the calibration of the one below, which can compound systematic errors along the ladder. The field has measured $H_0$ since Hubble's 1929 paper \citep{1929PNAS...15..168H} and has disagreed about the value for almost as long: Sandage and de Vaucouleurs differed by a factor of two for twenty years, until the Hubble Space Telescope Key Project settled the value at $72 \pm 8$ km s$^{-1}$ Mpc$^{-1}$ \citep{2001ApJ...553...47F}.

The current disagreement, the Hubble tension, is the most precisely stated in this history, and its significance depends on the indicator. SH0ES, using Cepheids, finds $H_0 \approx 73$ km s$^{-1}$ Mpc$^{-1}$ and claim a disagreement beyond the 5$\sigma$ discovery threshold; JWST photometry indicates that stellar crowding does not bias those Cepheids \citep{2022ApJ...934L...7R, 2024ApJ...962L..17R}. A parallel JWST program replacing Cepheids with tip-of-the-red-giant-branch (TRGB) and J-region asymptotic giant branch (JAGB) carbon-star indicatorsreturns roughly 68 to 70 \citep{2025ApJ...985..203F}, and CMB-anchored $\Lambda$CDM gives 67--68 \citep{2020A&A...641A...6P}. The most precise local measurements remain above the CMB value, but the spread among local indicators is now comparable to the gap itself. That points to the rungs as the likeliest origin, though new physics is not excluded.

The one route that skips the ladder is geometric. Water megamaser disks orbiting the nuclei of nearby galaxies yield distances from their rotation curves and centripetal accelerations, and the six galaxies of the Megamaser Cosmology Project give $H_0 = 73.9 \pm 3.0$ km s$^{-1}$ Mpc$^{-1}$ \citep{2020ApJ...891L...1P}. Pushing that sample deeper into the Hubble flow is limited first by the sensitivity needed to detect the faint high-velocity maser features. The next-generation Very Large Array (ngVLA) will enable this.

The Habitable Worlds Observatory (HWO) would shorten the ladder rather than refine a rung. With sufficient calibration, if its resolution and sensitivity reach individual Cepheids, TRGB stars, and JAGB stars at distances of order 100 Mpc, deep in the Hubble flow, the supernova rung can be removed: geometric anchors would calibrate stellar tracers already in the Hubble flow.

\begin{quote}
\textit{What could NASA do?} The cosmic distance ladder sets the scale of the universe, but three rungs of systematics may limit its accuracy. If HWO reaches individual Cepheids, TRGB stars, and JAGB stars at $\sim$100 Mpc, it can replace the three-rung supernova ladder with a two-rung geometric-to-stellar measurement of $H_0$.
\end{quote}

\medskip
\noindent\textbf{Understanding Type Ia Standard Candles.} Both preceding measurements rest on the same object: Type Ia supernovae carry the Hubble diagram and form the third rung of the distance ladder, yet their standardization is only empirical and possible evolution of the population with cosmic time is a leading systematic error for both uses. The optical light curve is reprocessed nuclear gamma rays from the decay of $^{56}$Ni. The MeV lines themselves are more informative: their fluxes give the nickel mass that sets the luminosity, and their profiles give the ejecta mass, velocity, and geometry. That constrains progenitor and explosion models, including the single- versus double-degenerate question Astro2020 called a critical unknown, and tests whether explosion variations drive the diversity of optical light curves \citep{2021pdaa.book.....N}. Only SN 2014J has been measured this way, by INTEGRAL \citep{2014Natur.512..406C, 2014Sci...345.1162D}. The Compton Spectrometer and Imager (COSI), launching in 2027, will reach only the nearest events. Characterizing the population requires an order-of-magnitude improvement in MeV line sensitivity, tens of supernovae per year.

\begin{quote}
\textit{What could NASA do?} A MeV mission an order of magnitude more sensitive than COSI would measure $^{56}$Ni/$^{56}$Co lines for tens of Type Ia supernovae per year, opening a path to physical rather than empirical standardization of these critical standard candles.
\end{quote}

\medskip
\noindent\textbf{Standard Sirens.} The gravitational waveform of a merging compact binary determines its luminosity distance: General Relativity relates the amplitude and frequency evolution to the distance directly, with no standardization and no ladder \citep{1986Natur.323..310S}. For a neutron-star merger with an electromagnetic counterpart, the measurement is essentially free of astrophysical systematics; the floor is the detectors' amplitude calibration. GW170817 demonstrated the method at roughly 15\% precision on $H_0$ and remains the only such event \citep{2017Natur.551...85A}.

A precision program requires the third-generation detectors planned for the 2030s, Cosmic Explorer and the Einstein Telescope \citep{subcommittee2024report}, which would observe most binary neutron-star mergers in the observable universe. The waveform does not contain the redshift; that must come from electromagnetic observation, and gravitational-wave localizations alone are too coarse to identify a host galaxy. Nearby, wide-field optical surveys, such as Rubin's, recover kilonovae. At higher redshift the counterparts are short gamma-ray bursts and synchrotron afterglows, which can extend the siren Hubble diagram to $z\sim 2$--$3$, through the era of dark-energy domination \citep{2024EPJC...84..663H, 2026EPJC...86....8H}. That channel needs prompt detection and localization of the burst, then a redshift of the host or afterglow. The localization accuracy must be typically better than 10' to enable follow-up by most telescopes, or itself achieve arcsecond for robust host galaxy assocation. Forecasts from this counterpart sample reach a fraction of a percent on $H_0$ and percent-level constraints on $w$, independent of supernovae, BAO, and the distance ladder \citep{2026EPJC...86....8H}. Siren distances are calibrated by General Relativity, so a well-measured nearby sample would set the absolute scale for extragalactic astrophysics.

\begin{quote}
\textit{What could NASA do?} Standard sirens can measure the expansion history of the universe independently of supernovae, BAO, and the distance ladder. The enabling electromagnetic capability is more sensitive all-sky keV--MeV coverage, to detect and precisely localize ($\leq10'$) the short gamma-ray bursts that extend the diagram to $z\sim 2$--$3$.
\end{quote}

The Laser Interferometer Space Antenna (LISA), launching in 2035 with NASA contribution, adds an additional channel: massive black-hole mergers, seen to $z\gtrsim 8$. An accretion-powered counterpart would supply the redshift. The sample is small, and there is no guaranteed counterpart: identifying the specific galaxy, and thus redshift, requires prompt X-ray, optical, and radio follow-up over LISA's localization region, the last because a merger of two accreting black holes may launch jets \citep{2010Sci...329..927P} that very-long-baseline radio imaging could resolve. Weak lensing by intervening structure limits individual distances at high redshift \citep{2016JCAP...04..002T, 2025PhRvD.111h3043M}. Combined with the third-generation neutron-star sample, the siren Hubble diagram would span the expansion history from the local universe to $z\sim 7$, at the few-percent level at high redshift \citep{2025CQGra..42s5002S}.

\begin{quote}
\textit{What could NASA do?} LISA extends the siren Hubble diagram to the highest redshifts if prompt X-ray and optical/infrared follow-up can identify the counterparts of its massive black-hole mergers.
\end{quote}

\medskip
\noindent\textbf{Redshift Drift.} \citealt{1962ApJ...136..319S} showed that acceleration would appear as a slow change in the redshift of a single source, $\dot{z} = (1+z)H_0 - H(z)$, and estimated that a detection would take of order ten million years of monitoring. Decades later, \citet{1998ApJ...499L.111L} showed that the Lyman-$\alpha$ forest toward the brightest quasars, observed with stable spectrographs on large telescopes would require only decades. The shift is 1--2~cm/s per decade. A twenty-year campaign at $z=2$--$5$ is planned for the ANDES spectrograph on ESO's Extremely Large Telescope; ESPRESSO at the VLT is already taking the first epochs \citep{2008MNRAS.386.1192L}.

A decihertz gravitational-wave observatory sees $\dot{z}$ as a phase delay. A neutron-star binary that enters the band at 0.1~Hz remains in view for years; at $z=1$, acceleration delays the waveform by about 1~second over a ten-year observation \citep{2001PhRvL..87v1103S, 2012JCAP...04..031Y}. The same waveform supplies the luminosity distance. An ensemble of order a million binaries would then map distance and $\dot{z}$ without electromagnetic redshifts, enough to distinguish acceleration from inhomogeneous alternatives \citep{2012PhRvD..85d4047N}.

\begin{quote}
\textit{What could NASA do?} A decihertz observatory tracks neutron star and black hole binaries for years before merger whose measurement of redshift drift and luminosity distance provide a gravitational wave-only Hubble diagram. This would enable a unique probe of dark energy, independent of the cosmological ladder and light.
\end{quote}

\section{Q3: How Did the Universe Begin?}

A century of cosmology has traced the universe back to a hot, dense state. What produced that state is now an experimental question: a beginning, or an earlier epoch, perhaps a bounce. Inflation is the leading answer. Its predictions can be tested, and those tests reach energies about twelve orders of magnitude above colliders, near the scale where the fundamental forces may have been unified.

The standard chronology begins in a state of extreme density and temperature. Inflation is a hypothesized episode within the first fraction of a second in which space expanded exponentially, driven by the energy of a field called the inflaton \citep{1980ApJ...241L..59K,1981MNRAS.195..467S,1981PhRvD..23..347G}. The episode explains why the observable universe is flat and nearly uniform. It also leaves a testable signature: quantum fluctuations of the field were stretched to astronomical size, causing slight variations in density. Inflation ended when the field reached the bottom of its potential. Its energy was then converted into ordinary particles, a process called reheating, and the familiar hot expansion took over. After 380,000 years the plasma became transparent, releasing the light we observe today as the cosmic microwave background (CMB). Roughly a hundred million years with no luminous sources followed, the dark ages, in which gravity slowly amplified those density variations. The first stars ended the era, and over the next several hundred million years starlight reionized the gas between galaxies. The beginning itself has to be studied from still earlier epochs: the dark ages, the CMB, and inflation.

A dedicated CMB polarization mission was the first of Q2C's three new initiatives in 2003 \citep{council2003connecting}. The Roadmap placed that mission in its Formative Era, and an ultra-long wavelength radio array and a decihertz gravitational-wave observatory in its Visionary Era \citep{2014arXiv1401.3741K}. No document has yet treated the three as one program. The DETF did that for dark energy's independent probes; the same structure could be used here.

\medskip\noindent\textbf{The Cosmic Microwave Background.} The CMB records the primordial fluctuations in enough detail to test inflation indirectly. The power spectrum, which characterizes the strength of fluctuations and how it varies with scale, is parameterized by the amplitude $A_s$ and the spectral tilt $n_s$. In slow-roll inflation, where the inflaton descends a nearly flat potential gradually \citep{1982PhLB..108..389L, 1982PhRvL..48.1220A}, $n_s$ falls slightly below 1. A single slowly rolling field produces nearly Gaussian fluctuations, so any non-Gaussian correlation, measured by the parameter $f_{\rm NL}$, must satisfy $|f_{\rm NL}| \ll 1$. The same stretching applied to spacetime produces a companion background of primordial gravitational waves. Its strength relative to the density fluctuations, the tensor-to-scalar ratio $r$, measures the energy scale of inflation, and the models that best fit the measured tilt predict $r$ roughly between $10^{-3}$ and $10^{-2}$. Its most accessible imprint is a faint curl pattern in the CMB polarization, called B modes.

Inflation's rivals make their own predictions for the same observables. In loop quantum cosmology, an approach to quantum gravity, the Big Bang is not a beginning but a bounce from a prior contracting phase, and the bounce leaves potentially observable imprints on the fluctuations at the largest angular scales \citep{agullo2023loop}. Ekpyrotic and cyclic scenarios, in which the hot expansion follows a slow contraction, predict no observable primordial gravitational waves at all, so a detection of B modes would rule them out \citep{2004PhRvD..69l7302B}. The matter bounce scenario, in which primordial perturbations are generated during a matter-dominated contracting phase preceding a cosmological bounce, has also played an important role as an alternative to inflation \citep{finelli2002generation}. However, these models are generally already excluded by observation \citep{quintin2015evolution}. The future measurements to target then test whether our universe is the first, or bounced from a previous one.

NASA's COBE discovered the temperature anisotropies in 1992 \citep{1992ApJ...396L...1S}. WMAP mapped them across the full sky and measured $n_s$ below 1 \citep{2013ApJS..208...19H}, as the simplest models predict. Planck found the fluctuations consistent with Gaussian statistics, with no extra fields, and spatially flat once baryon acoustic oscillation data are included \citep{2020A&A...641A..10P}. BICEP/Keck has pushed the gravitational-wave limit to $r < 0.036$ \citep{2021PhRvL.127o1301A}, excluding the steepest inflaton potentials. Recent ground data have shifted $n_s$ slightly; the precise value is not yet settled.

Astro2020 identified sensitivity at the $10^{-3}$ level, the bottom of the favored range, as decisive either way: a detection would measure the energy scale of inflation, and a null result would rule out the leading models \citep{2021pdaa.book.....N}. A measurement of $r$ above 0.01 would further imply that the inflaton moved a distance larger than the Planck scale in field space, sharply constraining quantum-gravity theories. On the statistical side, a local $f_{\rm NL}$ of order $1$ would rule out single-field slow-roll inflation and/or point to new physics in the pre-inflationary era \citep{agullo2011non}. SPHEREx, NASA's all-sky infrared survey, is forecast to approach that $f_{\rm NL}$ level by the end of the decade \citep{2014arXiv1412.4872D}.

The other instruments now pursuing these thresholds are on the ground or abroad. The Simons Observatory, with its funded expansion, forecasts $\sigma(r) \approx 1.2 \times 10^{-3}$ \citep{2019JCAP...02..056A, 2026JCAP...04..051A} enough to detect the upper part of the favored range but not to rule out the leading models on a null result. CMB-S4, designed to do so and recommended by Astro2020 and the 2014 and 2023 P5 reports, was canceled in 2025, with no funded successor \citep{2016arXiv161002743A}. 

JAXA's LiteBIRD is designed for a $\sigma(r) < 0.002$ from the full sky, with launch targeted for 2036 and decision gates still ahead \citep{aizawa2026litebird}. The NASA line for the same measurement runs from Q2C's initiative through Beyond Einstein's Inflation Probe \citep{nasa2003beyond}, Astro2010's technology endorsement \citep{2010nwnh.book.....N}, the Roadmap's CMB Polarization Surveyor \citep{2014arXiv1401.3741K}, and Astro2020's suggestion of a CMB mission as a future Probe \citep{2021pdaa.book.....N}.

\begin{quote}
\textit{What could NASA do?} A NASA CMB probe is the remaining U.S. path to $r\sim 10^{-3}$ on the full sky, and to the large-scale polarization measurements that funded ground experiments cannot guarantee.
\end{quote}

\medskip\noindent\textbf{The Dark Ages.} Astro2020 named ``finding innovative ways to probe cosmology in the `dark ages' prior to any significant star formation'' one of its discovery areas \citep{2021pdaa.book.....N}. The dark ages are the roughly one hundred million years between the release of the CMB and the first stars, when the universe was neutral gas, dark matter, and light only from the CMB. Neutral hydrogen has a single spectral feature, the 21 cm hyperfine line, stretched by expansion to ultra-long radio wavelengths. That signal contains inflation information inaccessible in the CMB: the primordial power spectrum on several more decades of spatial scale \citep{2004PhRvL..92u1301L}, and a local $f_{\rm NL}$ reach well beyond Planck if foregrounds can be controlled \citep{bull2024modes}. 

Late in the dark ages, at redshifts of 30 to 200, the line lies at 7 to 50 MHz. Earth is a poor site: the ionosphere distorts the signal and, at the low end, blocks it, and terrestrial transmitters fill the band. The lunar far side is the only nearby site shielded from that interference, and during lunar night from the Sun's radio emission as well. The Roadmap placed a far-side radio array in its Visionary Era \citep{2014arXiv1401.3741K}. The first NASA-DOE step is the Lunar Surface Electromagnetics Experiment--Night (LuSEE-Night), a pathfinder landing in early 2027 to characterize the far-side radio environment \citep{2023arXiv230110345B}. NASA-funded concept studies have mapped the path from single stations to a lunar array.

\begin{quote}
\textit{What could NASA do?} The ultra-long wavelength radio dark-ages 21 cm signal is the only known electromagnetic window between recombination and the first stars. It is accessible only from the lunar far side, and the needed sensitivity requires substantial technological innovation.
\end{quote}

\medskip\noindent\textbf{Decihertz Gravitational Waves.} No light reaches us from before recombination: photons scattered in the primordial plasma until the CMB was released. Only two known messengers reach us from earlier times. Relic neutrinos decoupled about one second after the Big Bang and have never been directly detected. Gravitational waves travel unimpeded from the moment they are generated, so a primordial background would give a direct view of the first instants, and its amplitude would measure the energy scale of inflation. The CMB samples that background at one effective frequency, near $10^{-17}$ Hz. A space detector in the decihertz band, near 0.1 Hz, would sample the same spectrum roughly sixteen orders of magnitude higher, giving data close to creation, at wavelengths that re-entered the horizon about $10^{-18}$ seconds after the Big Bang. Two samples this far apart constrain the spectrum's shape, a stronger discriminator among inflation models and their rivals than either amplitude alone. If reheating, which filled the universe with heat and particles, was still under way when these wavelengths re-entered, its imprint sits in this band as well, a feature no other path presently reaches.

Beyond Einstein named a Big Bang Observer, built on Laser Interferometer Space Antenna (LISA) heritage, as a vision mission, and the Roadmap placed a $\sim$0.1 Hz gravitational-wave mapper in its Visionary Era \citep{nasa2003beyond,2014arXiv1401.3741K}. No decihertz space detector is in formal development by any agency.

\begin{quote}
\textit{What could NASA do?} A decihertz detector would sample the primordial gravitational-wave spectrum sixteen orders of magnitude above the CMB. The expected amplitude depends on $r$, which the CMB program will measure or bound. The mission would require technological innovation beyond LISA.
\end{quote}

\section{Q4: Did Einstein Have the Last Word on Gravity?}

Is General Relativity the correct theory of gravity, or does it break down somewhere: near the horizons of black holes, in the dynamical spacetime of a merger, on the cosmological scales where it requires dark matter and dark energy to fit the data, or at the Planck scale, where a quantum theory must take over? At the Planck scale it must, though its effects may not be observable. Gravity stands apart from the other fundamental forces: the electromagnetic, weak, and strong interactions are quantum theories described within the Standard Model, while gravity remains classical geometry. Naive attempts to quantize it fail. A quantum theory of gravity is among the deepest open problems in physics, and a measured deviation from General Relativity would provide a path forward. For its first century \citep{1916AnP...354..769E} the theory was tested in the weak-field regime and then in the slow-motion strong-field regime. The quarter-century since Q2C has opened the dynamical strong field.

The classical Solar System tests, the deflection of starlight by the Sun \citep{1920RSPTA.220..291D}, the perihelion advance of Mercury, and the Shapiro delay of radar signals passing near the Sun \citep{1964PhRvL..13..789S}, confirmed the theory to ever-higher precision \citep{2003Natur.425..374B}. The orbital decay of the Hulse-Taylor binary pulsar provided the first indirect evidence that gravitational waves exist \citep{1975ApJ...195L..51H,1982ApJ...253..908T}, and NASA's Gravity Probe B directly measured the geodetic and frame-dragging precessions of gyroscopes in Earth orbit \citep{2011PhRvL.106v1101E}. The classical tests now reach the center of the Milky Way: the star S0-2/S2, orbiting the four-million-solar-mass black hole Sgr~A* every sixteen years, shows the gravitational redshift at closest approach and the same relativistic orbital precession first measured at Mercury, in a gravitational potential ten thousand times deeper \citep{2018A&A...615L..15G, 2019Sci...365..664D, 2020A&A...636L...5G}. A pulsar in a tight orbit around Sgr~A* would go further. Timing its orbit would measure the black hole's spin and quadrupole moment, and combined with the stellar orbits and horizon-scale imaging would test the no-hair theorem \citep{2016ApJ...818..121P}. No ideal object has been found; identification would require advanced radio or X-ray surveys. 

In the theory's centennial year, LIGO detected the merger of two black holes, and the signal set the first gravitational-wave bounds on the mass of the graviton and on dispersion of the waves \citep{2016PhRvL.116f1102A, 2016PhRvL.116v1101A}. Two years later the gamma-ray burst of the neutron-star merger GW170817 arrived at NASA's Fermi observatory 1.7 seconds behind the gravitational waves, after both had traveled for 130 million years, limiting any difference between the speed of gravity and the speed of light to about one part in $10^{15}$ \citep{2017ApJ...848L..13A}. LIGO-Virgo-KAGRA has now observed nearly 400 mergers, and none has required a departure from the theory \citep{2026arXiv260527225T, 2026arXiv260719293T}. The Event Horizon Telescope (EHT) imaged the predicted black hole shadows of M87* and Sgr~A* on event-horizon scales \citep{2019ApJ...875L...1E, 2022ApJ...930L..12E}. General Relativity further predicts that astrophysical black holes are described by the Kerr solution \citep{1963PhRvL..11..237K} and specified by mass and spin alone, the no-hair theorem.  The loudest merger yet recorded, GW250114, resolved two ringdown tones of the newborn remnant, both consistent with the Kerr spectrum, and showed that the total horizon area grew through the merger, as Hawking's area theorem requires \citep{2025PhRvL.135k1403A}.

The remaining tests divide by frequency band and by scale. The high-frequency range of stellar-mass mergers belongs to the ground-based network: Cosmic Explorer, recommended in the MPSAC ngGW report \citep{subcommittee2024report}, and the European Einstein Telescope would extend current sensitivity roughly tenfold. Tests on cosmological scales, including the relation of gravity to dark energy, are carried by the survey facilities of Q2. Networks of next-generation atomic clocks, a topic raised in the BPS Decadal \citep{medicine2023thriving}, would sharpen weak-field redshift tests and, later, could open a mid-band gravitational-wave window between LISA and the ground.

The instrument for the millihertz band is the Laser Interferometer Space Antenna (LISA), conceived in the 1990s as a joint ESA-NASA mission. AANM ranked it second among moderate space missions \citep{2001aanm.book.....N}; Q2C reaffirmed it as a test of General Relativity at the boundary of physics and astronomy; Beyond Einstein made it one of two Einstein Great Observatories \citep{nasa2003beyond}; and Astro2010 ranked it third among the large space missions \citep{2010nwnh.book.....N}. NASA withdrew under budget pressure in 2011, and ESA rebuilt the mission at reduced scale; after the 2015 detection NASA returned as a partner, and Astro2020 recommended that NASA work with ESA to ensure LISA achieves the full capability envisioned a decade earlier \citep{2021pdaa.book.....N}. ESA formally adopted the mission in January 2024 for a 2035 launch, and NASA's hardware contribution is now a formal project.

LISA will observe massive black-hole mergers at signal-to-noise ratios in the thousands. It will also track extreme mass-ratio inspirals (EMRIs): a stellar-mass object completing of order $10^{5}$ orbits deep in the field of a massive black hole, mapping the multipoles of the surrounding spacetime \citep{2024arXiv240207571C}. That is among the most stringent tests of the Kerr metric and the no-hair theorem foreseeable this century.

\begin{quote}
\textit{What could NASA do?} Ensure the success of the LISA mission, which enables the most stringent foreseeable tests of the Kerr metric and the no-hair theorem.
\end{quote}


Light that completes extra half-orbits around a black hole before escaping forms a stack of subrings, each exponentially narrower and closer to a critical curve set by the mass, spin, and inclination \citep{2020SciA....6.1310J, 2020PhRvD.102l4004G}. The ring the EHT measures is the direct emission, whose size depends on where the plasma radiates; that calibration against simulations is why the M87* and Sgr~A* consistency tests sit near ten percent. The first subring tracks the critical curve more closely. Its interferometric diameter follows the curve to a few percent, with the residual set by the geometry of the emitting gas \citep{2023PhRvD.108f4043C}. Averaging over multiple epochs allows the separation of this ring, being a property of spacetime, and the fluctuations occurring in the plasma. However, for these two black holes, measurement of this signal requires baselines longer than are possible on the surface of Earth.

Explorer class missions have been proposed which could work with the ground telescopes which could measure the first subrings of M87* and Sgr~A* \citep{2024SPIE13092E..2DJ}, which can probe the Kerr metric at a few percent accuracy \citep{2022A&A...657L..12G}. While LISA's EMRI is a more powerful test of the no-hair theorem, these observations would probe spacetime with light instead of gravity, and these two black holes are prototypes for many astrophysics studies. The second subring is a null Kerr test below the percent level, insensitive to the emitting gas and valid at inclinations to roughly 45 degrees. Measurement of this would require baselines roughly those from Earth to the Moon \citep{2020PhRvD.102l4004G, 2022A&A...668A..11P}. 

\begin{quote}
\textit{What could NASA do?} A space millimeter-interferometry would measure the photon rings of M87* and Sgr~A*, probing spacetime and providing precise spin measurements. At the Galactic Center it would sit beside the stellar orbits and a pulsar, if one is found, as a third constraint on mass, spin, and quadrupole. Submillimeter stations at lunar distance would later reach the second subring, a sub-percent test that does not depend on the emitting gas.
\end{quote}

\section{Q5: What Are the Masses of the Neutrinos, and How Have They Shaped the Evolution of the Universe?}
When Q2C posed the neutrino-mass question in 2003 \citep{council2003connecting}, neutrino oscillations were only a few years old: neutrinos have mass, and the Standard Model as originally written says they should not. The two halves of the question are closely linked, because the same free-streaming physics by which neutrinos influenced the growth of cosmic structure makes cosmology the most sensitive measure of their mass. 

\medskip
\noindent\textbf{The Laboratory Frontier.} For most of the twentieth century the neutrino was assumed massless, like the photon. Neutrino oscillations overturned that assumption: a neutrino created in one flavor (electron, muon, or tau) can be detected later as another, which is possible only if the flavor states are quantum superpositions of states of definite mass ($\nu_1$, $\nu_2$, $\nu_3$) and at least two of those masses differ from zero. This was established by the 1998 atmospheric result at Super-Kamiokande and the 2001 solar result at SNO \citep{1998PhRvL..81.1562F, 2001PhRvL..87g1301A}. Neutrino mass remains the one proven flaw in the minimal Standard Model: as originally written, the theory gives the neutrino only left-handed states and no mechanism to acquire mass.

What oscillation experiments measure is not the masses but the differences of their squares. Two splittings are now known precisely \citep{2020JHEP...09..178E}, and they fix the spacing of the three mass eigenstates but not their absolute values. The remaining unknowns are the absolute scale, the ordering (whether $\nu_3$ is the heaviest, called normal, or the lightest, called inverted), and whether the neutrino is its own antiparticle (Majorana) or not (Dirac). A measurement of the sum of the three masses, combined with the splittings, determines the individual masses. The splittings set a floor the sum must exceed, 0.059 eV for the normal ordering and 0.10 eV for the inverted. These are experimentally constrained by the kinematic endpoint of tritium beta decay, which is model independent; neutrinoless double-beta decay, nonzero only if neutrinos are Majorana; and cosmology, sensitive to the sum through the imprint of neutrinos on the growth of structure.

If neutrinos are their own antiparticles, a seesaw with heavy right-handed partners can explain why the light masses are small \citep{1977PhLB...67..421M}. Related heavy states can drive leptogenesis \citep{1986PhLB..174...45F}; a separate keV-scale sterile neutrino is a dark-matter candidate. Neutrinoless double-beta decay would establish lepton number violation and that neutrinos are Majorana, which is why the search has been a top construction priority of the 2015 and 2023 NSAC LRPs \citep{committee2015reaching, committee2023new}.

Two decades of reactor and accelerator experiments have refined the splittings and mixings; the third mixing angle was measured in 2012 \citep{2012PhRvL.108q1803A}, and the long-baseline program has produced first indications, not yet conclusive, of leptonic charge-parity (CP) violation \citep{2020Natur.580..339T}, a difference between the oscillations of neutrinos and antineutrinos whose observation would support neutrino-sector explanations of the cosmic matter excess. KATRIN has set the strongest model-independent kinematic limit \citep{2025Sci...388..180K}. The first generation of large double-beta-decay experiments has pushed half-life limits into the inverted-ordering band \citep{abe2025search}. JUNO has begun its mass-ordering campaign \citep{2026Natur.654..343J}.

The Deep Underground Neutrino Experiment (DUNE) and Hyper-Kamiokande will determine the mass ordering and measure the CP-violating phase \citep{ritz2014building, 2024arXiv240719176A}. The ton-scale double-beta-decay campaign, the highest-priority new construction of the 2023 NSAC LRP \citep{committee2023new}, is designed to cover the inverted-ordering band, and will be underground. The absolute mass scale is the exception: for that quantity, the most sensitive measurement comes from the sky.

\medskip
\noindent\textbf{The Cosmological Frontier.} Neutrinos are the second most abundant particle in the universe, as there are a few hundred relic neutrinos in every cubic centimeter, and they have influenced cosmic history at every epoch \citep{2006PhR...429..307L}. In the first second they helped set the neutron-to-proton ratio that fixed the primordial helium and deuterium yields of Big Bang Nucleosynthesis. Through the structure-forming era the massive relics acted as a small hot component of the dark matter, streaming out of small-scale potential wells and suppressing the clustering of matter below their free-streaming scale, a few-percent effect for a mass sum near the oscillation floor; the same physics ruled out the neutrino-dominated hot-dark-matter cosmology decades ago. If leptogenesis is correct, the cosmic matter excess originates in the neutrino sector. Because the mass signal is this same structure-suppression signal, the two halves of Q5 are answered by the same measurements.

Cosmological data now constrain the mass sum roughly an order of magnitude more tightly than the laboratory, and the tightest limits sit near or below the oscillation floor. The Dark Energy Spectroscopic Instrument (DESI) combined with the cosmic microwave background (CMB), under $\Lambda$CDM, gives $\Sigma m_\nu < 0.064$ eV \citep{elbers2025constraints}, a central value consistent with zero. The tension depends on the model: it relaxes with the newer Planck likelihood, shifts with the supernova compilation, and largely disappears if $w$ is allowed to evolve or if $\tau$ is substantially higher than Planck's value \citep{2025arXiv250716589D, 2026arXiv260630903S}. The cosmological inference is still entangled with two other quantities, the dark energy equation of state $w(z)$ and the optical depth to reionization $\tau$. Progress on Q5 requires resolving both. The neutrino signal is a suppression of growth, and evolving dark energy can mimic it. Separating the two requires measuring expansion and growth together. DESI, Spec-S5, Rubin, Euclid, and Roman are designed to do that \citep{2015arXiv150303757S}.

The remaining limitation is $\tau$, the fraction of CMB photons rescattered after reionization. It is degenerate with the mass sum in every CMB-anchored analysis, and its present uncertainty caps the cosmological measurement near $\sigma(\Sigma m_\nu) = 0.020$ eV regardless of survey quality \citep{2019JCAP...01..059B}. The cosmic-variance limit is $\sigma(\tau) = 0.002$, and reaching it requires mapping CMB polarization on the largest angular scales over the full sky, which requires observation from space. With that measurement, the survey program reaches $\sigma(\Sigma m_\nu) \approx 0.012$ eV, sufficient for a 5$\sigma$ detection of the minimal mass and separation of the orderings; if the central value instead remains at zero, the result is a $>4\sigma$ conflict with the oscillation experiments \citep{2024JHEP...09..097C}. LiteBIRD is designed to deliver this measurement \citep{2023PTEP.2023d2F01L}, but it is not assured: the mission is in concept development, with launch targeted for 2036 and decision gates still ahead.

No other funded project provides this measurement. CMB-S4 was canceled in 2025; remaining ground experiments supply late-time lensing but not a cosmic-variance $\tau$ \citep{2024arXiv240719176A, 2021pdaa.book.....N}. The large-scale E modes themselves are pursued from the ground by CLASS and from a balloon by Taurus, but neither can reach the requirement: their best forecasts saturate near $\sigma(\tau)=0.003$--$0.004$, above the cosmic-variance limit, bounded by sky fraction and foregrounds rather than by detectors \citep{2018ApJ...863..121W, 2020apra.prop...49B, 2022ApJ...940...68E}. Without a cosmic-variance $\tau$, DESI, Rubin, Euclid, and Roman cannot reach a detection of the oscillation floor. The requirement is the measurement, on whichever platform provides it.

\begin{quote}
\textit{What could NASA do?} Ensure that a CMB cosmic-variance-limited measurement of the reionization optical depth is made, the single number that limits the cosmological determination of the neutrino masses. Without it, Roman and the coming surveys remain capped near $\sigma(\Sigma m_\nu)=0.020$ eV, short of the oscillation floor.
\end{quote}

An ultra long wavelength radio 21~cm array measures the same free-streaming suppression, in a still-linear spectrum, before stars form, and with far more independent modes than the CMB. The neutrino signature is a feature in the shape of $P(k)$ at each redshift, not in the overall amplitude, so it is independent of $\tau$. At the oscillation-floor mass, the heaviest eigenstate becomes non-relativistic near $z\sim 100$, inside the dark ages 21~cm window, and a redshift series would see the suppression appear. Forecasts for a cosmic-variance-limited array approach meV precision on $\Sigma m_\nu$, an order of magnitude past the survey-program target \citep{2004PhRvL..92u1301L, 2008PhRvD..78f5009P}. 

\begin{quote}
\textit{What could NASA do?} An ultra-long wavelength radio 21 cm array would ultimately deliver a precise measurement of the neutrino masses, an order of magnitude beyond that expected from the planned survey programs.
\end{quote}

\section{Q6: How Do Cosmic Accelerators Work and What Are They Accelerating?}

The most energetic particles ever detected are cosmic rays exceeding $10^{20}$ eV, about the kinetic energy of a tennis ball struck in a rally, carried by a single atomic nucleus. The origin of these particles has been an open question since their discovery more than sixty years ago. Q2C posed it in 2003 in two parts, asking how the accelerators work and what they are accelerating \citep{council2003connecting}.

When an accelerator drives protons and nuclei to extreme energies, their collisions with surrounding gas and radiation produce secondary particles whose decays yield gamma rays and neutrinos; a single hadronic source therefore emits all three messengers, which do not carry the same information. Cosmic rays arrive deflected by magnetic fields and do not point back to their sources. Gamma rays travel straight but are ambiguous, since accelerated electrons alone can radiate them; a gamma-ray source need not accelerate nuclei at all. Neutrinos are the clean signature, produced only in collisions of accelerated protons and nuclei and arriving undeflected. Q2C set out to find the origins of all three and supported a major facility for each, which became NASA's Fermi gamma-ray observatory, the IceCube neutrino detector in the Antarctic ice, and the Pierre Auger cosmic-ray observatory in Argentina \citep{council2003connecting}. 

\medskip
\noindent\textbf{Particle Astrophysics.} IceCube discovered a diffuse flux of high-energy astrophysical neutrinos in 2013, establishing that the universe contains accelerators of matter reaching the TeV--PeV range \citep{2013Sci...342E...1I}. In 2023 it detected the Milky Way in neutrinos, a diffuse glow consistent with cosmic rays interacting with interstellar gas, and the first view of our Galaxy in a messenger other than light \citep{2023Sci...380.1338I}. Among individual sources, the first candidate, a blazar flaring in 2017, remains debated \citep{2018Sci...361.1378I}. A Galactic candidate has also appeared. The X-ray binary Cyg X-3, whose jet drives into the dense wind of its Wolf-Rayet companion, showed its most closely associated IceCube events during its brief GeV flaring periods \citep{2023MNRAS.524L..89K}, with additional evidence from flare timing \citep{2025arXiv251216638T,2023MNRAS.524L..89K}. Such an association can be confirmed or rejected with continued radio, X-ray, and gamma-ray monitoring. The strongest association is with the nearby active galaxy NGC 1068, which produces TeV neutrinos without the accompanying high-energy gamma rays that an optically thin hadronic source would produce \citep{2022Sci...378..538I, 2025ApJ...981..131A}. The deficit requires that the neutrinos be produced in a compact region opaque to gamma rays; if they are produced from photohadronic interactions they should be within roughly 100 Schwarzschild radii of the accreting supermassive black hole \citep{2022ApJ...941L..17M}. The favored site is the hot X-ray corona above the accretion disk; winds and jet-cloud interactions remain alternatives, while the starburst and large-scale jet, both transparent to gamma rays, cannot supply the signal \citep{2020PhRvL.125a1101M, 2020ApJ...891L..33I, 2022ApJ...941L..17M}. Excesses reported from additional Seyfert galaxies point to a neutrino source population, and such hidden cores may supply much of the astrophysical neutrino flux below a few tens of TeV \citep{2025ApJ...988..141A, 2020PhRvL.125a1101M}. IceCube-Gen2 would raise the detection rate roughly tenfold and test the hidden-core hypothesis against X-ray-bright active galaxies \citep{2021JPhG...48f0501A}. It is endorsed by Astro2020 and the 2023 P5 report, and is not yet in construction due to Antarctic infrastructure problems \citep{2021pdaa.book.....N, 2024arXiv240719176A}.

The gamma-ray view is split between space and the ground. Fermi has surveyed the GeV sky for nearly two decades, establishing the gamma-ray populations of active galaxies and supernova remnants and detecting the pion-decay signature that shows they accelerate protons, though not to the highest energies \citep{2013Sci...339..807A}. LHAASO surveys from hundreds of GeV to PeV energies and has detected the most energetic photons ever recorded, identifying ``PeVatrons,'' accelerators reaching the energies where the Galactic cosmic-ray spectrum is expected to end \citep{2021Natur.594...33C}. LHAASO and HAWC have since found emission above 100 TeV from several microquasars, among them SS 433 and V4641 Sgr, making X-ray binary jets a Galactic accelerator class in their own right \citep{2024Natur.634..557A, 2025NSRev..12af496L}. The Cherenkov Telescope Array Observatory (CTAO), under construction with roughly ten times the sensitivity of current instruments, is the pointed complement to the survey instruments: it will resolve and characterize the gamma-ray signals that Fermi and LHAASO can detect but not sharply image \citep{2019scta.book.....C}. 

NASA's opportunity lies in the medium-energy gamma-ray band, from roughly 0.1 MeV to a GeV, closing the gap between the hard-X-ray telescopes and Fermi. The gamma rays absorbed alongside the neutrinos do not vanish: reprocessed through cascades, their power re-emerges where the source becomes transparent, and the models concentrate that escaping emission at 1--10 MeV, where the corona's thermal emission has died away, with a luminosity that tracks the neutrino luminosity \citep{2020PhRvL.125a1101M}. As no sufficiently sensitive telescope in this band has been launched, this prediction remains untested. The Compton Spectrometer and Imager (COSI), launching in 2027, will make the first sensitive measurements in that window, but only for a small sample of sources \citep{2024icrc.confE.745T}.

At higher energies, because the opacity to pair production climbs with energy, each source's spectrum must turn over between roughly 1 MeV and a GeV, at an energy set by the compactness of the production region, and measuring that turnover constrains the size of the neutrino production region \citep{2022ApJ...941L..17M}. Separately, neutral-pion decay imprints a spectral feature near 70 MeV that leptonic emission cannot easily mimic, the definitive signature of accelerated protons, which Fermi could extract for only the two brightest supernova remnants at the edge of its band \citep{2013Sci...339..807A}. Coverage from tens of MeV through the pion peak near a GeV would extend that test across the accelerator population, including the PeVatron candidates LHAASO is identifying. Both measurements demand angular resolution as much as sensitivity in order to avoid source confusion.

\begin{quote}
\textit{What could NASA do?} A medium-energy gamma-ray survey from roughly 0.1 MeV to a GeV, with sensitivity and angular resolution well beyond COSI and Fermi, to: test the predicted 1--10 MeV counterparts of hidden neutrino sources, measure the pair-production turnover that locates the production region, and extend the pion-decay test across the Galactic accelerator population. COSI's observations, beginning in 2027, will clarify the expectations around which such a mission would be designed or exclude this approach.
\end{quote}

The cosmic rays are the accelerated particles themselves, and at the highest energies their study belongs to ground-based observatories. The Pierre Auger Observatory and the Telescope Array traced the spectrum to its end, finding the suppression long expected from interactions with the cosmic microwave background near $5\times10^{19}$ electron-volts, though whether those interactions or the accelerators reaching their limiting energy produces the cutoff is not yet known \citep{2020PhRvL.125l1106A}. A large-scale anisotropy in arrival directions places the sources outside the Galaxy \citep{2017Sci...357.1266P}. The two arrays also disagree: Auger's data favor a composition growing heavier toward the highest energies, while the Telescope Array's smaller sample remains consistent with protons \citep{2014PhRvD..90l2005A, 2023Sci...382..903T}. Whether this is real or a cross-calibration artifact is unresolved, which is why measuring the mass of each particle has become a key goal. AugerPrime is already operating for this purpose \citep{2016arXiv160403637T}. A next-generation ground observatory, the Global Cosmic Ray Observatory, has been proposed for the 2030s \citep{2023arXiv230917324A}.

A detector in orbit would view the highest-energy air showers from above, enabling uniform exposure over the full sky. If the particles are light, this can eliminate the cross-calibration problem at the energies above about 20 EeV where magnetic deflection is smallest. A mission concept based on flown balloon and space-station pathfinders, the ultra-high energy (UHE) concept, would provide this measurement and could search for tau neutrinos skimming Earth's limb \citep{2021JCAP...06..007P}. It is the space-based heir to the Auger recommendation with which Q2C began.

\begin{quote}
\textit{What could NASA do?} Observe the highest-energy cosmic rays and Earth-skimming tau neutrinos from orbit, with uniform full-sky exposure at the energies where magnetic deflection is smallest.
\end{quote}

\medskip
\noindent\textbf{Matter and Energy Near Black Holes.} For the active galaxies suggested as the origin of high-energy neutrinos the engine is a supermassive black hole and its accretion flow. What is not understood is the precise engine mechanism that accelerates the particles, which also affects jet composition including whether the emission is hadronic or leptonic in origin. A jet can be launched powered by the energy in the disk, Blandford--Payne, or from field lines threading the spinning hole, Blandford--Znajek, which taps the hole's rotational energy \citep{1982MNRAS.199..883B, 1977MNRAS.179..433B}. Confirming the second would show that the ergosphere is being tapped, a strong-gravity result (Q4).

Jet power in Blandford--Znajek scales as the square of the horizon flux and the square of the horizon angular velocity \citep{1977MNRAS.179..433B}. The Event Horizon Telescope's 2021 polarized image of M87* resolved an ordered, dynamically important magnetic field at the jet base, and Faraday rotation measures the flux and the accretion rate that sustains it \citep{2021ApJ...910L..12E, 2021ApJ...910L..13E}. Magnetically arrested simulations reach jet efficiencies above unity, drawing power that must be supplied by the black hole's rotation \citep{2011MNRAS.418L..79T}. Spin is the remaining relevant parameter, but it cannot currently be measured for jetted nuclei. Radiatively inefficient flows lack the disk reflection features that X-ray spectroscopy fits; claimed correlations between X-ray-binary spins and radio flares remain contested \citep{2012MNRAS.419L..69N, 2013MNRAS.431..405R}; and a spin taken from fits of horizon-scale images to the same simulations that assume the mechanism cannot test it. The photon ring (Q4) gives the spin from lensing geometry. A low spin under a jet of $10^{42}$--$10^{45}$~erg~s$^{-1}$ would rule out spin extraction in M87*; a high spin at the measured accretion rate would leave the black hole's rotation as the term that can cover the power.

The Earth caps ground arrays near four gravitational radii for M87*; an orbiting antenna reaches 1.5, enough to separate horizon-anchored field lines from disk-anchored ones, and at 3~mm, where the jet base is bright, it would extend the ring--jet connection the ground array has already traced \citep{2023Natur.616..686L}. AANM recommended an orbiting antenna, ARISE, for the jet-launch region \citep{2001aanm.book.....N}. It was not built. Russia's RadioAstron flew the centimeter-wave version from 2011 to 2019 and found jet cores above $10^{13}$~K, past the inverse-Compton limit and still unexplained \citep{2016ApJ...820L...9K}. The millimeter band would repeat that measurement without interstellar scattering.  The Roadmap assigned space interferometry in radio or X-rays as a Visionary Era instrument \citep{2014arXiv1401.3741K}.

\begin{quote}
\textit{What could NASA do?} Fly a millimeter antenna with the ground array. It would measure the spin of M87* and image the jet to the launch point, testing whether that jet draws power from the hole's rotation. X-ray spectroscopy does not reach the radiatively inefficient, jetted population; the Roadmap's X-ray interferometer is the later alternative.
\end{quote}

Timescales associated with jet launching, and probes of jet acceleration, can come from multi-wavelength timing measurements, which reach spatial scales not accessible to direct imaging with current or near-future facilities.  In X-ray binaries, lags of seconds to minutes of jet emission in the infrared through radio bands, behind X-ray emission from the accretion disk, have been observed in a few cases (e.g. \citealt{2021MNRAS.504.3862T,2024Natur.627..763R}), and support models in which the jets are launched at speeds close to the escape velocities from the launch region \citep{2024Natur.627..763R} and powered by internal shocks caused by modest variations in the jet speeds \citep{2018MNRAS.480.2054M}.  Qualitatively consistent results have been seen in long-timescale monitoring of the active galactic nucleus (AGN) NGC~7213 \citep{2011MNRAS.411..402B}.  Such work presently consists of snapshots; no source has intensive monitoring of the time lags and correlation coefficients across bands.  For the X-ray binaries, one needs an X-ray timing instrument with flexible scheduling and good coordination with ground-based facilities, to collect many such measurements across objects and accretion rates and learn how jet launching and acceleration depend on accretion rate, compact-object spin, and the black hole or neutron star nature of the accretor.  For AGN, coordinated long-timescale monitoring campaigns are needed, but need not be strictly simultaneous.

Substantial efforts have been made to estimate spins using reflection spectroscopy with facilities like XMM-Newton and NuSTAR \citep{2014ApJ...782...76G,2025ApJ...989..227D}.  In X-ray binaries, where alternative techniques like disk continuum spectroscopy are also accessible (e.g. \citealt{2016ApJ...825...13S}), several cases of poor agreement exist, and some carry substantial model dependence even within one method \citep{2024ApJ...962..101Z}.  Fitted inner disk inclination angles often disagree with binary orbital inclination angles; frame dragging can cause this, but the implied non-planar disks would present large outer-disk solid angles, producing strong narrow emission line components not incorporated in the modeling \citep{2021ApJ...906...28A}.

Spectroscopy alone is thus unlikely to solve the black hole spin problem, although after suitable calibration of the models, it may.  Calibration is best done using X-ray timing, which measures the light travel times between where the continuum is produced and where it illuminates the geometrically thin accretion disk.  The NICER mission has measured these lags well enough to estimate the size scale of the hard X-ray emission region \citep{2022ApJ...930...18W}; the next step is to measure lags independently for different energies within the broad iron line.  With the lags encoding size scales, and the energies encoding Doppler shifts and gravitational redshifts, degeneracies can be broken and the geometry and ionization state of the disk understood in detail \citep{2000ApJ...529..101Y}.  This approach can give a true understanding of black hole spins, but requires a timing mission with collecting area at Fe~K$\alpha$ far larger than NICER's.  With these insights in hand, one can address the model systematics in time-averaged reflection spectroscopy, applicable to far more systems with missions like NewAthena.

The neutrino results add a target. The corona implicated in NGC 1068's neutrinos emits predominantly in X-rays and, given the lack of MeV coverage, its X-ray luminosity is used to predict the neutrino output \citep{2020PhRvL.125a1101M}. Its temperature and structure are measured only through X-ray spectroscopy, timing, and polarization, and the winds XRISM resolves are the leading alternative production site \citep{2022ApJ...941L..17M}. IXPE has made the first such polarization measurement, in NGC 4151, finding the corona extended in the plane of the disk rather than compact on its axis \citep{2023MNRAS.523.4468G}. A flagship-class observatory would characterize coronae across the accreting population and resolve the nearest sources.

The near-term goal is a study for a flagship-class X-ray observatory, a capability that has stood in NASA's strategic plans for a quarter-century. Gas spiraling toward a black hole emits a relativistic iron line whose shape encodes the black hole's spin and the structure of the flow feeding it, and X-ray spectroscopy is the only technique that can measure spin across the population of luminous accreting black holes. That was the promise of Constellation-X, ranked second only to JWST among large space missions in AANM and endorsed by Q2C as a probe of the spacetime just outside the horizon \citep{2001aanm.book.....N, council2003connecting}; the Roadmap carried it forward as the X-Ray Surveyor \citep{2014arXiv1401.3741K}. JAXA's XRISM, with a NASA-supplied spectrometer, is now resolving the nuclear winds of active galaxies into their component velocities \citep{2025Natur.641.1132X}. 

\begin{quote}
\textit{What could NASA do?} Study a next-generation X-ray observatory: spin spectroscopy across the accreting population, hard-X-ray coverage and polarimetry of the coronae that may be IceCube's hidden accelerators, timing of possible pulsars orbiting Sgr~A*, and the collecting area to resolve nearby engines. A flagship study would rank those capabilities.  An X-ray facility capable of large numbers of coordinated observations with ground-based facilities is needed to understand jet launching via disk-jet time lag measurements, and could also calibrate spin spectroscopy models. Longer term, the Roadmap's X-ray interferometer would image the jet-launching region.
\end{quote}

\medskip
\noindent\textbf{New Windows on the Dynamic Universe.} A fourth messenger arrived after Q2C and after the 2013 Roadmap. LIGO's first gravitational-wave detection in 2015 and the neutron-star merger GW170817 in 2017 opened a channel unlike the other three: cosmic rays, gamma rays, and neutrinos are products of an accelerator's surroundings, while gravitational waves arise from the motion of the central masses themselves \citep{2016PhRvL.116f1102A, 2017ApJ...848L..12A}. Astro2020 recast Q2C's origins question around the sources rather than the particles: New Windows on the Dynamic Universe \citep{2021pdaa.book.....N}.

Much of this science now arrives as transients and depends on detecting them promptly at wavelengths accessible only from space. A neutrino or gravitational-wave detection becomes most powerful with an electromagnetic counterpart, which for neutron-star mergers with future gravitational wave detectors is most readily supplied by gamma-ray monitors. The detailed needs for such an instrument are covered in the report from the Gamma-ray Transient Network Science Analysis Group \citep{burns2023gamma}. Astro2020 put that coordinated time-domain and multimessenger astrophysics capability first among sustaining space activities.

\begin{quote}
\textit{What could NASA do?} Establish a strategic time-domain and multimessenger fleet: more sensitive all-sky coverage of the high-energy sky, with rapid imaging, spectroscopic, and polarimetric follow-up in X-ray, ultraviolet, and far-infrared. Such facilities need to be capable of providing deep characterization, not only detection.
\end{quote}

\section{Q7: Are Protons Unstable?}

No proton has ever been observed to decay. The proton is stable in the Standard Model, but only by accident: baryon number is conserved because every interaction among the Standard Model's particles at ordinary energies leaves it unchanged. The proton, being the lightest particle that carries baryon number, has no allowed decay. In contrast, Grand Unified Theories (GUTs) allow decay. They merge the strong, weak, and electromagnetic interactions at a high scale, put quarks and leptons in the same multiplets, and let superheavy gauge bosons turn one into the other. For that gauge-boson process the lifetime grows as the fourth power of the unification mass. An observed proton decay would be evidence beyond the Standard Model and would fix the unification scale.

The search for proton decay is also motivated by cosmology. The observable universe contains matter and essentially no antimatter, roughly six baryons for every ten billion photons, and Sakharov showed in 1967 that generating this excess from symmetric initial conditions requires three conditions, the first being a process that changes baryon number \citep{1967ZhPmR...5...32S}. The Standard Model's anomalous electroweak processes can redistribute such an asymmetry but cannot create one of the observed size, so baryon- or lepton-number violation beyond the Standard Model must have operated in the early universe. Unified theories provide it naturally, through the decays of the same superheavy bosons that mediate proton decay or through leptogenesis involving the heavy seesaw neutrinos discussed in Q5. Proton decay would be the low-energy remnant of that physics.

The experimental program began with the minimal SU(5) theory of \citet{1974PhRvL..32..438G}, whose predicted lifetime near $10^{30}$ years was excluded by the Irvine-Michigan-Brookhaven (IMB) and Kamiokande experiments in the 1980s \citep{1983PhRvL..51...27B, 1989PhLB..220..308H}. For three decades since, the most sensitive searches have come from Super-Kamiokande. When Q2C made determining the proton lifetime a central recommendation \citep{council2003connecting}, the lower limit in the benchmark channel $p\rightarrow e^{+}\pi^{0}$ stood near $5\times10^{33}$ years; Super-Kamiokande has since raised it to $2.4\times10^{34}$ years \citep{2020PhRvD.102k2011T}. The corresponding limit in the channel favored by supersymmetric unification, $p\rightarrow\bar{\nu}K^{+}$, is $5.9\times10^{33}$ years \citep{2014PhRvD..90g2005A}. Gauge-boson-mediated decay points to $e^{+}\pi^{0}$ lifetimes of order $10^{35}$ years. Supersymmetric unification favors $\bar{\nu}K^{+}$, and the LHC exclusion of superpartners up to multi-TeV masses pushes those predictions into the $10^{34}$--$10^{36}$-year range. For the gauge-boson channel, each factor of 5 in lifetime reach is only a factor of 1.5 in unification scale.

Three next-generation facilities are now operating or in advanced construction. Hyper-Kamiokande, with more than 8 times Super-Kamiokande's sensitive mass, begins operations in 2028. The Deep Underground Neutrino Experiment (DUNE), a key priority of successive P5 reports, using liquid-argon chambers that image candidate decays track by track, expects data late this decade. JUNO, a liquid-scintillator detector that identifies kaons by their timing signature, has been taking data since 2025. Together they reach about $10^{35}$ years in the benchmark channel and roughly $3\times10^{34}$ years in the kaon channel \citep{2018arXiv180504163H, 2020arXiv200203005A, 2023ChPhC..47k3002A}. The three techniques are complementary, so the same mode can be cross-checked with more than one detector. By the late 2030s the program will either have seen proton decay or have closed much of the remaining GUT window. Nothing beyond this generation is approved, so that result is likely to stand for decades. Astrophysical observation cannot substitute for these detectors, being orders of magnitude short of the laboratory limits. Astro2020 contains no proton-decay science case \citep{2021pdaa.book.....N}. 

\begin{quote}
    \textit{What could NASA do?} Nothing.
\end{quote}

\section{Q8: What Are the New States of Matter at Exceedingly High Density and Temperature?}

Every substance has a phase diagram. For water it plots pressure against temperature, dividing the plane into ice, liquid, and vapor. The boundaries between them are phase transitions. The boiling transition is sharp and first-order, with two coexisting phases and latent heat, yet it ends at a critical point, beyond which liquid and vapor blend into one another. Locating a substance's transitions, and establishing which are sharp and which are smooth, reveals what it is made of and how its constituents interact. This question asks for the phase diagram of matter governed by the strong nuclear force \citep{council2003connecting}, described by quantum chromodynamics (QCD), the theory that binds quarks and gluons into protons and neutrons.

The strong force is weak when quarks are close and grows stronger as they separate, a property called asymptotic freedom \citep{1973PhRvL..30.1343G, 1973PhRvL..30.1346P}. Under ordinary conditions quarks are confined within protons and neutrons. At sufficiently high temperature or density, quarks are pressed so close together that the interaction weakens, and matter deconfines into a plasma of free quarks and gluons. The reference density is the nuclear saturation density $\rho_0$, the density inside a large atomic nucleus, so extreme that a liter of it would outweigh Mount Everest. First-principles theory covers only part of the diagram. Lattice QCD, which simulates the theory on a discretized grid of spacetime points, is reliable at high temperature but fails at high density. Controlled expansions of nuclear forces hold only to about twice $\rho_0$, short of neutron-star cores. The rest of the diagram has to be measured.

The diagram shares water's temperature axis, but its second axis is the baryon chemical potential $\mu_B$, a measure of the excess of matter over antimatter. The hot side filled the universe in its first microseconds and is recreated in heavy-ion collisions. The dense side exists today only in neutron stars, cold in their cores and briefly hot when they merge.

\medskip
\noindent\textbf{The Hot Side.} The hot side is the domain of the nuclear-physics community, which identified matter at extreme energy density as a frontier in the 2003 X-Games and 2004 NSTC HEDP reports \citep{council2003frontiers, nstc2004frontiers} and has built and steered the accelerator program through successive NSAC LRPs \citep{committee2015reaching, committee2023new}. A quarter-century of heavy-ion collisions at the Relativistic Heavy Ion Collider (RHIC) and the Large Hadron Collider (LHC) has shown that the quark-gluon plasma exists and that it flows as a near-perfect fluid, with viscosity near a conjectured quantum lower bound \citep{2005PhRvL..94k1601K}. The plasma gives way to ordinary matter through a smooth crossover, at a temperature that lattice QCD has pinned to percent-level precision \citep{2019PhLB..795...15B}. RHIC's Beam Energy Scan then carried the measurements into the baryon-rich region, where the crossover is predicted to steepen into a first-order transition ending at a critical point, but it has not yet been found. RHIC has ended operations. The Electron-Ion Collider (EIC), to be built partly in RHIC's tunnel, begins in the mid-2030s and will measure how quarks and gluons build nuclei, the starting point of the dense-matter problem.

\medskip
\noindent\textbf{The Dense Side.} The cold, dense extreme cannot be created in any laboratory. Probing it requires observations of neutron stars, which compress more than the mass of the Sun into the size of a city and reach several times $\rho_0$ in their cores. No single measurement spans the relevant densities. The 2023 NSAC LRP therefore organizes the effort as an equation of state density ladder, in analogy with the cosmological distance ladder, each rung constraining the same physics over an overlapping range of densities \citep{committee2023new}. The astrophysical rungs have stood in the planning record for a quarter-century. AANM endorsed Constellation-X \citep{2001aanm.book.....N}. The Roadmap named the nuclear equation of state as a goal, to be pursued through pulse-profile modeling with the Neutron Star Interior Composition Explorer (NICER) and through spectroscopy of redshifted lines from the stellar surface \citep{2014arXiv1401.3741K}. Astro2020 then elevated such science in its New Windows on the Dynamic Universe science theme \citep{2021pdaa.book.....N}. 

The bridge between observation and microphysics is the equation of state, the relation between pressure and density. Through the Tolman-Oppenheimer-Volkoff equations of relativistic stellar structure \citep{1939PhRv...55..364T, 1939PhRv...55..374O}, it maps onto what can be measured: a neutron star's mass and radius, its tidal deformability, meaning how readily a companion's gravity distorts it, and the maximum mass $M_{\rm TOV}$, above which no neutron star can resist collapse to a black hole. A stiffer equation of state, one generating more pressure at a given density, supports larger stars and a higher maximum mass; a softer one yields more compact stars and a lower one. If every neutron star is made of the same cold catalyzed matter, each precise measurement constrains the same function. Dark matter bound inside the stars would break that assumption (Q1). Different observables reach different rungs. 

The lowest rung is anchored in the laboratory. PREX-II and CREX at Jefferson Lab measured the neutron skins of lead-208 and calcium-48, the excess layer of neutrons at each nucleus's surface \citep{2021PhRvL.126q2502A, 2022PhRvL.129d2501A}. The measurements constrain the symmetry energy, the energy cost of a neutron excess, at saturation density. The Facility for Rare Isotope Beams (FRIB) produces the neutron-rich nuclei that tighten the constraint. Its proposed FRIB400 energy upgrade would reach toward twice saturation density, the same region the gravitational-wave rung probes from above \citep{frib2023frib400}.

The next rungs currently come from X-ray observations of the neutron stars themselves. NICER, on the ISS, pioneered pulse-profile modeling of millisecond pulsars, yielding mass and radius together for a few pulsars at roughly the ten-percent level \citep{2019ApJ...887L..21R, 2025ApJ...995...60M}. The enhanced X-ray Timing and Polarimetry mission (eXTP), led by the Chinese Academy of Sciences and approved for launch in 2030, carries pulse-profile modeling forward with modestly larger collecting area \citep{2025SCPMA..6819502Z}. NewAthena, with its NASA contribution, pursues both timing and spectroscopy and forecasts mass and radius at the few-percent level per source \citep{2025NatAs...9...36C}. However, each has limitations preventing ideal dense matter studies.  In particular, studies of burst oscillations in X-ray binaries \citep{2018IAUS..337..209W} are likely to be needed to access the higher mass neutron stars, which are crucial, as they are the objects whose radii also cannot be well-estimated using gravitational waves.

\begin{quote}
\textit{What could NASA do?} Deliver the NASA contribution to Athena. Support a high-collecting area timing mission that can provide pulse profile modeling of a larger sample of neutron stars, with broader properties, than Athena can. 
\end{quote}

The high-mass end is anchored by pulsar timing, which sets the lower bound on $M_{\rm TOV}$. In the rare, nearly edge-on binaries PSR J1614$-$2230 and PSR J0740+6620, the companion's gravity delays the pulsar's ticks by a measurable amount, the Shapiro delay. That delay weighed both stars at about two solar masses \citep{2010Natur.467.1081D, 2021ApJ...915L..12F}. Any viable equation of state must therefore support at least that mass, ruling out the simplest models softened by hyperons (baryons containing strange quarks) or free quarks. Heavier pulsars, if securely weighed, would raise that bound. Tensions between rungs are themselves constraints: the thick neutron skin from PREX-II and the smaller radii favored by some NICER analyses pull the symmetry energy in opposite directions.

Neutron star mergers enable several constraints at once. The late inspiral encodes the tidal deformability, as GW170817 showed, constraining the equation of state near twice $\rho_0$ \citep{2018PhRvL.121p1101A}. In a merger of a neutron star with a black hole, whether the neutron star is torn apart before it is swallowed, and so whether any light appears, depends on its radius. In a merger of two neutron stars, whether the remnant collapses promptly to a black hole or briefly survives depends on $M_{\rm TOV}$, and the two outcomes may leave distinct electromagnetic signatures. The post-merger signal, kilohertz gravitational waves from the hot remnant, would probe matter above nuclear density at temperatures of tens of MeV, in the hot and dense corner. It sits above the most sensitive band of Cosmic Explorer and the Einstein Telescope, but the few events loudest enough for the measurement are all that is needed. 

\begin{quote}
\textit{What could NASA do?} Build the coherent time-domain and multimessenger fleet. Continuous, far more sensitive all-sky gamma-ray monitoring would catch and localize the mergers. Rapid X-ray, ultraviolet, optical, and infrared follow-up would then characterize each counterpart while it is bright. The counterparts constrain the equation of state in a density and temperature range no laboratory can reach.
\end{quote}

While gamma-ray burst monitors offer immediately, precise localizations of neutron star mergers, such instruments do not detect the vast majority of mergers which are not face-on. A decihertz interferometer would detect binaries years before merger, providing precise measurements of intrinsic properties and providing arcminute scale localizations days in advance \citep{2020CQGra..37u5011A}. These precise, gravitational-wave only localizations, enable concrete determination of whether binaries produced light or not, enabling firm determination of the maximum mass of a neutron star. Multiwavelength gravitational wave astronomy with decihertz and kilohertz detections of neutron star mergers improves the measurement of tidal deformability by a factor of about 4 \citep{2018PTEP.2018g3E01I}. Thus, while the decihertz band provides unique information that operate together with other facilities to probe dense matter.

\begin{quote}
\textit{What could NASA do?} Develop a space-based decihertz gravitational-wave interferometer. Advance localization would let the electromagnetic fleet observe the fate of nearby mergers from onset, and the years-long inspiral records would improve the tidal deformability measurements of the ground-based gravitational wave detectors.
\end{quote}

\section{Q9: Are There Additional Space-Time Dimensions?}

Additional spatial dimensions were proposed by Kaluza and Klein in the 1920s as a route to unifying gravity with electromagnetism \citep{1921SPAW.......966K,1926ZPhy...37..895K}. They are required in string theory, but not in all quantum gravity approaches, so detection or exclusion is a path towards grand unification. In the theory's conventional form the extra dimensions are compactified near the Planck length of $10^{-33}$ cm, far beyond any direct experimental reach. The developments that provided a possible path forward, and led Q2C to pose the question \citep{council2003connecting}, came in the late 1990s.

Gravity is the one force without a quantum description, and many candidate theories endow spacetime itself with structure at the Planck scale. If that structure exists, light crossing cosmological distances can reveal it through an energy-dependent speed of light or a slow rotation of polarization, although this constrains extra dimensions only in models that predict altered photon propagation \citep{1999A&A...345L..32K}. Fermi's timing of gamma-ray burst photons constrains any linear energy dependence of the speed of light to scales beyond the Planck energy \citep{2009Natur.462..331A}.

The joint arrival of gravitational waves and gamma rays from GW170817 fixed the speed of gravity to that of light to one part in $10^{15}$ \citep{2017ApJ...848L..13A}, eliminating broad classes of theories built to mimic dark energy and dark matter through modified gravity. The relative arrival times, and the comparison of the gravitational-wave distance with the host distance, test models in which gravity can take different paths or lose amplitude into extra dimensions \citep{2018PhRvD..97f4039V,2018JCAP...07..048P}. Lorentz invariance has survived every astrophysical test. Future precision alone does not motivate a dedicated mission; these bounds will keep improving as a byproduct of missions built for other reasons.

Arkani-Hamed, Dimopoulos, and Dvali proposed in 1998 that extra dimensions could be large, up to a fraction of a millimeter, with gravity's weakness explained by dilution into the extra volume \citep{1998PhLB..429..263A}; Randall and Sundrum proposed in 1999 a warped extra dimension whose Kaluza--Klein modes sit at the TeV scale, not a large extra volume \citep{1999PhRvL..83.3370R}. Large extra dimensions predicted deviations from Newton's inverse-square law at submillimeter separations and, at the Large Hadron Collider (LHC), the production of Kaluza-Klein gravitons and, in some scenarios, microscopic black holes. The ATLAS and CMS experiments found neither; the null results set lower limits on the fundamental gravitational scale between roughly 6 and 11 TeV, depending on the number of dimensions \citep{2021PhRvD.103k2006A}. The High-Luminosity LHC will extend the mass reach by roughly 30 percent by the mid-2030s, a modest gain against a parameter space that extends well beyond collider reach. For the smallest numbers of extra dimensions the strongest limits are astrophysical: Kaluza-Klein graviton emission would have altered the cooling of SN 1987A and heated old neutron stars, and those archival limits reach the thousand-TeV range \citep{2003PhRvD..67l5008H}.

Laboratory tests of gravity provide the complementary constraint. Torsion-balance experiments have verified the inverse-square law down to separations of 52 $\mu$m, limiting a single extra dimension to about 30 $\mu$m \citep{2020PhRvL.124j1101L,2020PhRvL.124e1301T}, already below the roughly 85 $\mu$m length scale set by the dark-energy density \citep{2007PhRvL..98b1101K}. Compact dimensions are also generically accompanied by light scalar fields whose exchange would violate the universality of free fall; the French satellite mission MICROSCOPE found no violation at one part in $10^{15}$, a hundredfold improvement on ground-based tests \citep{2022PhRvL.129l1102T}. A swampland argument (the ``dark dimension'') puts a single extra dimension near a micron, set by the dark-energy scale \citep{2023JHEP...02..022M}. The coming decade's laboratory experiments target separations of 1--10 $\mu$m, covering much of that range.

\citet{2000PhLB..485..208D} allowed an infinite extra dimension, with gravity leaking off our 3-brane at large distances, as an alternative to dark energy. The Roadmap listed ``dimensions of space-time beyond the four of our everyday experience'' among the speculative goals of its millihertz gravitational-wave mission \citep{2014arXiv1401.3741K}. 

The self-accelerating proposal suffers from a ghost instability \citep{2007CQGra..24R.231K} and is excluded as the cause of acceleration \citep{2008PhRvD..78j3509F}. The gravitational-wave test it prompted survives. A wave that leaks into extra volume makes the waveform distance larger than the electromagnetic distance. The binary neutron-star merger GW170817, at 40 Mpc, gave $D = 4.02^{+0.07}_{-0.10}$ \citep{2018JCAP...07..048P}. The Laser Interferometer Space Antenna (LISA), launching in 2035, will extend the damping test by two orders of magnitude in distance through massive-black-hole binary mergers, reaching the gigaparsec crossover scales that leakage would need to imitate dark energy \citep{2022PhRvD.105f4061C}. The damping test cannot reach a compact micron-scale dimension, whose Kaluza-Klein excitations lie seventeen orders of magnitude above millihertz graviton energies; the gravitational-wave and laboratory programs therefore probe disjoint corners of the question.

Beyond LISA, the discovery path runs through laboratory gravity at micron separations, collider searches, and equivalence-principle tests. A micron-scale laboratory detection would change how a cosmological-constant dark energy is interpreted. Astro2020 does not discuss extra dimensions \citep{2021pdaa.book.....N}.


\section{Q10: How Were the Elements from Iron to Uranium Made?}

The origin of the chemical elements is among the oldest questions in astrophysics, beginning with the recognition in the 1920s that stars shine by fusing light elements into heavier ones. It took its modern form in 1957, when Burbidge, Burbidge, Fowler, and Hoyle \citep{1957RvMP...29..547B} and, independently, Cameron \citep{1957PASP...69..201C} classified the nuclear processes that build the periodic table. The broad picture they established still stands: hydrogen and most helium were made in the Big Bang, the elements up to the iron peak are produced by fusion in stars and their explosions, and almost everything heavier than iron is built by the capture of neutrons onto seed nuclei. Neutron capture comes in several forms, a slow process (the s-process), in which neutrons are added gradually enough that unstable nuclei decay between captures, a rapid process (the r-process), in which neutrons are added so fast that nuclei are driven far from stability before they can decay, and an intermediate process (the i-process), which lies between the former two processes.

\medskip
\noindent\textbf{The Origin of the Heaviest Elements.} Q2C posed the question in 2003 as ``How were the elements from iron to uranium made?'' \citep{council2003connecting}. The r-process makes roughly half the elements beyond iron and all of the actinides, thorium and uranium among them, but it requires an extremely high density of free neutrons. The site where it occurs remained unidentified for six decades. In the past quarter-century two r-process sites have been identified, neutron-star mergers and giant flares from magnetars, the most strongly magnetized neutron stars. Additional long-standing candidates like common supernova have been excluded or disfavored.

Two kinds of sources were long considered for the r-process: the common core-collapse supernova, frequent enough to enrich the Galaxy steadily, and the much rarer neutron-star merger. Multiple independent lines of evidence indicate rare, high-yield events as r-process sites. Surveys of the most metal-poor halo stars show a wide star-to-star scatter in their heavy-element abundances; frequent enrichment would have homogenized these abundances, so the scatter indicates a rare source \citep{2021RvMP...93a5002C}. The ultra-faint dwarf galaxy Reticulum II records a single ancient r-process event far too large for any ordinary supernova, while most similar dwarfs show no enrichment at all \citep{2016Natur.531..610J}. Deep-sea crusts record recent nearby supernova activity through live $^{60}$Fe, yet they hold far less $^{244}$Pu than those supernovae would have delivered had they made the actinides \citep{2021Sci...372..742W}. Galactic chemical-evolution models add an independent argument. The growth of r-process abundances over Galactic time, and the scatter among the oldest stars, appear difficult to reproduce with neutron-star mergers alone, because their long delay times leave the earliest generations under-enriched. This may indicate that at least one additional channel acted early in the history of the Milky Way \citep{2019ApJ...875..106C}.

Major facilities have been constructed to answer this question. The 2007 and 2015 NSAC LRPs recommended construction of the Facility for Rare Isotope Beams (FRIB) to measure properties of neutron-rich nuclei \citep{committee2007frontiers, committee2015reaching}. Astro2010 anticipated Advanced LIGO and its worldwide detector network operating by the middle of the 2010s \citep{2010nwnh.book.....N}, with nearby neutron-star mergers the prime anticipated source. In 2017, roughly seventy facilities observed the binary neutron-star merger GW170817 and its kilonova AT2017gfo, a transient powered by the radioactive decay of newly synthesized heavy elements \citep{2017ApJ...848L..12A}. These observations established neutron star mergers as the first confirmed r-process site. Astro2020 then made time-domain and multimessenger astrophysics (TDAMM) its highest-priority sustaining activity in space \citep{2021pdaa.book.....N}.

Observations since have tested the remaining candidates and added a second site. Long gamma-ray bursts (GRBs) arise from the collapse of rapidly rotating massive stars, a debated candidate site for the r-process. Deep James Webb Space Telescope (JWST) observations of the exceptionally bright GRB 221009A revealed an ordinary gamma-ray burst supernova with no r-process signature \citep{2024NatAs...8..774B}. GRB 230307A, a long gamma-ray burst apparently produced by a merger, was followed into the infrared by JWST, yielding a tentative identification of tellurium \citep{2024Natur.626..737L}. The specific progenitors of such systems are debated. In 2025, a delayed gamma-ray component of the 2004 giant flare of the magnetar SGR 1806$-$20 was identified as decay photons from newly synthesized r-process nuclei, establishing magnetar giant flares as the second confirmed site \citep{2025ApJ...984L..29P}. While two sites have been identified, work remains to identify or exclude others, and to determine their relative contributions over the history of the Milky Way.

Discovering and characterizing these transient events is a coordinated, multi-facility undertaking. Although GW170817 was found through the gravitational-wave network, it remains the only kilonova discovered that way. All other kilonovae have been found in the follow-up of gamma-ray bursts, making wide-field gamma-ray monitors the most productive discovery channel. The Rubin Observatory and the Argus Array, an all-sky optical monitor now under construction, are expected to discover kilonovae in the nearby universe at optical wavelengths \citep{2022PASP..134c5003L}. Magnetar giant flares can be identified only by their prompt gamma-ray signature. Once an event is located, characterization proceeds through rapid and then deep follow-up. The timing requirement is most severe for a giant flare, which is predicted to produce a brief optical-ultraviolet transient, a scaled-down kilonova peaking within ten to fifteen minutes. Observing this transient would confirm heavy-element production in flares too distant for the delayed radioactive decay gamma-ray signal to be detected.

\begin{quote}
\textit{What could NASA do?} Construct a coherent time-domain and multimessenger astrophysics program, as outlined in Astro2020. Discovery requires continuous all-sky gamma-ray monitoring to find neutron-star mergers and magnetar giant flares, and rapid X-ray, ultraviolet, and infrared follow-up.
\end{quote}

A gravitational-wave detector observes a binary before it merges, so alerts can be issued in advance of the explosion. LIGO can detect a GW170817-like event at most a minute before merger, an interval set by the detectors' low-frequency cutoff. The proposed third-generation ground observatories, Cosmic Explorer and the Einstein Telescope, could roughly localize a nearby merger hours in advance, with greater precision closer to merger \citep{2018PhRvD..97l3014C}. A space-based interferometer in the decihertz band, between the ground detectors and the Laser Interferometer Space Antenna (LISA), could follow a binary for days or longer and localize it to arcminutes \citep{2020CQGra..37u5011A}. 

\begin{quote}
\textit{What could NASA do?} Develop a space-based decihertz gravitational-wave interferometer, occupying the band between LISA and the ground-based detectors. By detecting a neutron-star binary days before it merges and localizing it precisely in advance, such a mission would allow observation of the merger from onset across the electromagnetic spectrum. That coverage would separate the emission components and improve the r-process yield measured from each event.
\end{quote}

The most detailed elemental measurements come from time-resolved spectroscopy of the closest events, demonstrated by the campaign on GW170817. Because the ejecta expand and cool rapidly, the elements recombine through successive ionization states, and each recombination stage produces spectral features that appear and evolve over hours to days \citep{sneppen2024emergence}. The timing and evolution of these features give the identities of the elements and the velocity structure of the ejecta. Confident line detections exist in AT2017gfo and GRB 230307A, but assigning them to specific elements is uncertain, because the lines of the lanthanides and actinides, in the ionization states a kilonova passes through, are not yet known well enough for unique identification \citep{2019Natur.574..497W}. A laboratory and theoretical effort is under way to resolve this limitation. The Habitable Worlds Observatory (HWO), given rapid-response commanding, would capture the earliest, hottest phase in the ultraviolet and optical, when the most highly ionized species recombine. JWST and the ground-based Extremely Large Telescopes would then follow the later epochs into the infrared. Together these facilities would measure the full atomic-line evolution of a kilonova for the first time.

A complementary approach uses ancient stars. The oldest, most metal-poor stars of the Galactic halo and nearby dwarf galaxies each preserve the yield of essentially a single early r-process event, and chemical-evolution models use the retained abundance patterns to count the contributing sites. Many r-process elements, including those at the diagnostic abundance peaks, have usable lines only in the ultraviolet. Adding the ultraviolet raises the number of measurable r-process elements in a star by more than half and enables complete abundance mapping of the heaviest elements. Hubble has a nearly complete UV r-process pattern for one star. HWO could extend that measurement to of order one hundred stars \citep{2025arXiv250703180R}, enough to test the transient and chemical-evolution approaches against each other.

\begin{quote}
\textit{What could NASA do?} Give HWO rapid-response commanding for early ultraviolet and optical kilonova spectroscopy, and high-resolution ultraviolet spectroscopy of ancient r-process stars. Such capabilities may enable a complete answer on the Origin of the Elements after a century of work.
\end{quote}

\medskip
\noindent\textbf{The Origin of the Isotopes.} With the answer of the Origin of the Elements on the horizon, we can now focus on the successor goal: understanding the origin of isotopes. Nuclei are forged in extreme environments, and in many astrophysical explosions. The details of the progenitor and engine mechanisms affect the detailed isotopic abundances. Some of these isotopes are radioactive, and through decay emit gamma rays at characteristic energies, allowing measurement of the abundances of individual isotopes. The details of a given event can alter the isotopic abundances. Because MeV gamma rays interact only weakly with matter, they are among the cleanest diagnostics of the physics inside the explosion. The line intensity gives the mass of the isotope produced, and the Doppler shift and broadening give the velocity and geometry of the ejecta. The third TDAMM community workshop identified nuclear-line spectroscopy as one of the irreplaceable diagnostics connecting observations of cosmic explosions to the physics within them \citep{2025arXiv250203577B}.

Laboratory nuclear data, measured at facilities such as FRIB, its proposed upgrade, and the $N=126$ factory at Argonne National Laboratory, determine what each process can produce. Sustained theoretical work connects those nuclear inputs to astrophysical observables. Gamma-ray observatories then detect the isotopes and identify their sites. The search for the origin of the isotopes is therefore also a study of the physics of cosmic explosions, a key focus of Astro2020's time-domain and multimessenger program. Nuclear-line observations of SN 1987A and the young remnant Cassiopeia A confirmed the production of iron-group elements in supernovae and revealed strongly mixed, asymmetric ejecta, direct evidence for the convective engine now thought to drive core-collapse explosions \citep{1988Natur.331..416M, 2014Natur.506..339G}. Thermonuclear bursts on accreting neutron stars undergo nucleosynthesis through the rapid proton-capture process \citep{2001PhRvL..86.3471S}. In these events, the details of the explosion process depend on laboratory measurements, and are a key tie between nuclear and astrophysics. However, those that eject radioisotopes enable a unique diagnostic through the study of nuclear lines.

The near-term step is NASA's Compton Spectrometer and Imager (COSI), a wide-field gamma-ray spectrometer launching in 2027 \citep{2024icrc.confE.745T}. COSI will map the Galaxy in the lines of $^{26}$Al and $^{60}$Fe and in the 511-keV positron-annihilation line, survey for young supernova remnants in $^{44}$Ti, and observe the decay chain of $^{56}$Ni through $^{56}$Co in any sufficiently nearby Type Ia supernova. Mapping the origin of the isotopes across large samples of multiple transient populations will ultimately require a more sensitive successor. Such an instrument would enable deep characterization of Type Ia supernova and nova populations, studies of core-collapse supernovae, as well as observations of additional source classes.

\begin{quote}
\textit{What could NASA do?} Develop a next-generation nuclear-astrophysics, gamma-ray observatory as the successor to COSI. By resolving the decay lines of individual supernova remnants, novae, and explosive transients, it would determine where the isotopes are made and how cosmic explosions work.
\end{quote}

\section{Q11: Is a New Theory of Matter and Light Needed at the Highest Energies?}

When Q2C posed this question, the Higgs boson was undiscovered and the Large Hadron Collider (LHC) was under construction \citep{council2003connecting}. The theory at issue is quantum electrodynamics (QED) inside the Standard Model: whether they still hold at the greatest energies and field strengths nature reaches.

\medskip
\noindent\textbf{The Laboratory Frontier.} The Higgs boson was discovered by ATLAS and CMS at the LHC in 2012 \citep{2012PhLB..716....1A, 2012PhLB..716...30C}, completing the particle content of the Standard Model and validating the mechanism of electroweak symmetry breaking proposed almost fifty years earlier. Its couplings agree with Standard Model predictions, at roughly the 10\% level in most channels and a few percent in the best. Fifteen years of LHC searches have not found supersymmetric partners, new gauge bosons, extra Higgs bosons, compositeness, or large extra dimensions (Q9). The frameworks invented to explain the Higgs mass had predicted such states nearby. The Standard Model still has no accompanying new particles up to several TeV. Their absence has sharpened the hierarchy problem.

The muon magnetic moment, long the most prominent laboratory hint of new physics, is no longer a clear discrepancy. Fermilab's final result is now more precise than the theory it tests \citep{2025PhRvL.135j1802A}; a lattice-QCD evaluation of the hadronic contribution brings theory and experiment into agreement \citep{2025PhR..1143....1A}. What remains is a disagreement among theoretical inputs.

The High-Luminosity LHC begins physics around 2030, measuring Higgs couplings at the few-percent level, making the first real measurement of the Higgs self-coupling, and extending the mass reach for new particles. Beyond it, EPP-2024 \citep{medicine2025elementary}, following the 2014 and 2023 P5 reports \citep{ritz2014building, 2024arXiv240719176A}, recommends that the United States host the world's highest-energy collider around mid-century, and that it participate in the FCC-ee Higgs factory under study at CERN, to measure the Higgs couplings at the 0.1--1 percent level where new physics up to roughly 10 TeV would imprint itself indirectly.

Quantum electrodynamics is the most precisely tested sector in the Standard Model: its prediction of the electron's magnetic moment matches measurement to about one part in a trillion \citep{2023PhRvL.130g1801F}. Those tests are all in weak fields, where the vacuum is well-understood and dynamically unimportant. QED predicts that starting around $10\%$ the Schwinger critical field (electric or magnetic) \citep{1951PhRv...82..664S}, $B_{\rm QED} = m_e^2 c^3/(e\hbar) \approx 4.4\times10^{13}$~G (the strength at which the energy cyclotron energy in the field rivals its mass energy, $\hbar \omega_{\rm cyc} = m_e c^2$) the virtual electron-positron pairs that fill the vacuum become polarized and empty space turns into an active nonlinear optical medium: it acquires refractive indices that differ for the two polarizations of light (vacuum birefringence), it transmutes  photons below 1 MeV by splitting them in two (photon splitting) or if above 1 MeV converting them into real electron-positron pairs (or even neutrinos). Moreover, in a strong magnetic field, free electrons are constrained to discrete quantum mechanical states known as Landau states \citep{1966RvMP...38..626E,2006RPPh...69.2631H}.


That regime has never been created on Earth. There are two electromagnetic Lorentz invariants in QED: $\boldsymbol{E}\cdot \boldsymbol{B}$ and $E^2-B^2$. The distinction from terrestrial strong-field QED is essential. The strongest artificial magnetic fields ever produced, in single destructive laser pulses, fall short of $B_{\rm QED}$ by a factor of a million and are only sampled in a regime where $E^2-B^2 \approx 0$. Ultraintense lasers thus probe rapidly varying electromagnetic fields rather than a sustained, magnetic vacuum with Landau states. What the laboratory has verified is the underlying interaction: ATLAS has observed light scattering off light in heavy-ion collisions \citep{2019PhRvL.123e2001A}, and photon splitting has been claimed in the electric fields of heavy nuclei at the Budker Institute \citep{2002PhRvL..89f1802A}, both in agreement with QED. However, both are in a regime where the magnetic field is weak, and where Landau states are unimportant. Neither terrestrial experiment sends radiation through a macroscopic, near-critical magnetic field. The terrestrial collider fields are microscopic and transient, while the Budker experiment probes an atomic electric field.  The strong-field consequences remain beyond reach: dedicated searches for vacuum birefringence sit a factor of several below the QED prediction \citep{2020PhR...871....1E}. Thus the underlying QED interactions are tested, but the regime in which birefringence, magnetic photon splitting, magnetic pair creation, and Landau quantization operate together has not been reproduced on Earth. The strong-field, nonlinear regime is reached only in the sky.

\medskip
\noindent\textbf{The Astrophysical Frontier.} The untested QED predictions need supercritical magnetic fields, which in the present-day universe only magnetars provide. Their surface fields of $10^{14}$--$10^{15}$~G exceed $B_{\rm QED}$ by an order of magnitude \citep{2017ARA&A..55..261K}; ordinary pulsars at $10^{12}$--$10^{13}$~G fall short. Because the electric fields induced in a magnetar magnetosphere remain subcritical even where the magnetic field is far supercritical, the predictions on display are those of the supercritical magnetic vacuum: vacuum birefringence and photon splitting. 

Q2C flagged magnetized neutron stars as a place where new behavior of matter and light could appear. It did not become a named priority in later planning for astrophysics as Astro2020 and the 2013 Roadmap do not discuss strong-field QED \citep{2021pdaa.book.....N, 2014arXiv1401.3741K}. Progress has come from smaller missions.

In a magnetized vacuum the two polarization states of light, electric vector parallel or perpendicular to the field, travel at slightly different speeds. Around a magnetar the consequence is large and observable: as X-rays from the hot surface, or hard X-rays and soft gamma-rays from a burst, propagate outward, birefringence forces each ray's polarization to track the local field direction until far from the star, where the field direction is nearly uniform across the visible hemisphere, so radiation from different parts of the emitting region arrives aligned rather than mutually canceling \citep{2003MNRAS.342..134H,2026arXiv260522323W}. An inert vacuum generally predicts modest net polarization due to cancellation of polarization from an extended emitting region. In contrast,  a birefringent one predicts polarization far higher due to polarization reorientation. The underlying vacuum nonlinearity was predicted by Heisenberg and Euler in 1936 \citep{1936ZPhy...98..714H} and vacuum birefringence by \citet{1952PhDT........21T}.

IXPE found magnetars among the most strongly polarized celestial X-ray sources. Its 2022 observation of 4U 0142+61 showed a $90^\circ$ swing of polarization angle with energy, the first indication of QED mode physics \citep{2022Sci...378..646T}. The strongest case so far is the 2025 coordinated radio and X-ray polarimetric campaign on 1E 1547.0$-$5408: phase-averaged polarization reaches 65 percent at 2 keV, higher at some rotational phases approaching $100\%$, with the polarization direction locked to the magnetic field. Surface-emission modeling constrained by the radio-derived geometry cannot reproduce this without QED vacuum birefringence, a result arriving ninety years after the prediction \citep{stewart2026vacuum}. 

Continued IXPE campaigns extend the test across the magnetar population. GOSoX will open the 0.2--0.4 keV band below IXPE \citep{marshall2026new} where thermal surface emission dominates and model predictions diverge most sharply. Together they will confirm or refute the current claim. One band will then remain: the hard X-rays and soft gamma rays, where magnetar persistent emission is thought to arise from resonant scattering in the strongest-field regions of the magnetosphere \citep{2019BAAS...51c.292W} or quantum synchrotron emission \citep{2025ApJ...991..178H}, and where the vacuum's effect on light extends beyond a rotation of its polarization.

In a supercritical field, light itself can behave as a particle in a medium. A photon can split in two, a process forbidden in empty space, at a rate that climbs as the fifth power of the photon energy. Photon splitting can be regarded as ``photon Cherenkov radiation" \citep{1966RvMP...38..626E}. Above the pair-creation threshold near 1 MeV, photons can instead convert directly into real electrons and positrons. Crucially, photon splitting operates below 1 MeV where it has smoking-gun signatures in energy-dependent polarization.  Neither process has been seen in a magnetized vacuum. Their signatures lie in how magnetar emission behaves at its highest energies: cutoff energies set by the field along each line of sight \citep{2018ApJ...854...98W,2025ApJ...991..178H}. The cleaner signature is the polarization and its energy dependence around those cutoffs: in the weak-dispersion limit the QED selection rules allow only one of the two modes to split \citep{1971AnPhy..67..599A}, so the escaping radiation can be almost fully polarized, a pattern emission geometry alone is not expected to produce. It is actually not known how photon splitting behaves in the nonlinear strongly dispersive regime (several conversion modes are allowed by charge-parity symmetry). These ```smoking-gun" energy-dependent polarization signatures, and understanding the nonlinear QED vacuum, are highlighted in \citet{2019BAAS...51c.292W}.

The range of interest lies between tens of keV and a few MeV, the natural band of Compton telescopes. COSI, launching in 2027, surveys the 0.2--5 MeV band with high spectral resolution and provides a first look \citep{2024icrc.confE.745T}, but its effective area and polarization response are unlikely to yield a clean detection even in the brightest outbursts. A next-generation MeV mission with greater continuum sensitivity and polarimetric capability could deliver the first test of photon splitting in the persistent magnetar population, and possibly also bursts.

\begin{quote}
\textit{What could NASA do?} Deliver unique tests of quantum electrodynamics in its untested nonlinear regime, where strong magnetic fields turn the vacuum itself into an active optical medium. By observing magnetars, the only objects in nature whose fields reach this regime, a next-generation MeV mission with polarimetric capability would complete the confirmation of vacuum birefringence and provide the first tests of photon splitting.
\end{quote}

\bibliography{bibliography}

\end{document}